\documentclass[pra,aps,amsmath,twocolumn,letterpaper,superscriptaddress,showpacs]{revtex4}
\usepackage{bm,graphicx,amsmath}
\usepackage{bbm}

\usepackage{amssymb}
\usepackage{amsmath}
\usepackage{amstext}
\usepackage{latexsym}
\usepackage{dsfont}
\usepackage{color}

\newcommand{\ave}[1]{\left\langle#1\right\rangle}
\newcommand{\aven}[1]{\big\langle#1\big\rangle}

\newcommand{\tr}[1]{{\rm Tr}\{#1\}}

\newcommand{\XX}{X} 
\newcommand{\PP}{P} 
\newcommand{\RR}{R} 
\newcommand{\xx}{{\bf x}} 
\newcommand{\dxx}{\di{\bf x}} 
\newcommand{\pp}{{\bf p}} 
\newcommand{\dpp}{\di{\bf p}} 
\newcommand{\ee}{{\bf e}} 
\newcommand{\bgamma}{{\boldsymbol{\gamma}}} 
\newcommand{\bzeta}{{\boldsymbol{\zeta}}} 
\newcommand{\bxi}{{\boldsymbol{\xi}}} 

\newcommand{\tp}{{T}}
\newcommand{\totalVar}{\mathcal{V}}
\newcommand{\di}[1]{\bar{#1}} 
\newcommand{\basind}[1]{{\scriptscriptstyle(#1)}}
\newcommand{\omcut}{{\di\omega_{\rm cut}}} 
\newcommand{\op}[1]{{\rm #1}}
\newcommand{\crc}{{\basind{\pm}}}
\newcommand{\tth}{{\di t_{\rm th}}}
\newcommand{\trev}{{\di t_{\rm rev}}}
\newcommand{\DXsquare}{{\Delta\di X_+^2}}
\newcommand{\DPsquare}{{\Delta\di P_+^2}}

\newcommand{\Mdiff}{d} 

\newcommand{\diff}[2]{\frac{\Mdiff#1}{\Mdiff#2}}

\newcommand{\diffz}[2]{\frac{\Mdiff^2#1}{\Mdiff#2^2}}

\begin{document}

\title{Statistical mechanics of entanglement mediated by a thermal reservoir I}

\author{Endre Kajari}
\affiliation{Theoretische Physik, Universit{\"a}t des Saarlandes, D-66123 Saarbr{\"u}cken, Germany}
\author{Alexander Wolf}
\affiliation{Theoretische Physik, Universit{\"a}t des Saarlandes, D-66123 Saarbr{\"u}cken, Germany}
\author{Eric Lutz}
\affiliation{Department of Physics, University of Augsburg, D-86135 Augsburg, Germany}
\affiliation{Dahlem Center for Complex Quantum Systems, FU Berlin, D-14195 Berlin, Germany}
\author{Giovanna Morigi}
\affiliation{Theoretische Physik, Universit{\"a}t des Saarlandes, D-66123 Saarbr{\"u}cken, Germany}
\affiliation{Grup d'{\`O}ptica, Departament de F{\'i}sica, Universitat Aut{\`o}noma
de Barcelona, E-08193 Bellaterra, Spain}

\date{\today}

\begin{abstract} 
Two defect particles that couple to a harmonic chain, acting as common reservoir, can become entangled even when the two defects do not directly interact and the harmonic chain is effectively a thermal reservoir for each individual defect. This dynamics is encountered for sufficiently low temperatures of the chain and depends on the initial state of the two oscillators. In particular, when each defect is prepared in a squeezed state, entanglement can be found at time scales at which the steady state of a single defect is reached. We provide a microscopic description of the coupled quantum dynamics of chain and defects. By means of numerical simulations, we explore the parameter regimes for which entanglement is found under the specific assumption that both particles couple to the same ion of the chain. This model provides the microscopic setting where bath-induced entanglement can be observed.
\end{abstract} 


\maketitle

\section{Introduction}

It is commonly understood that the coupling of quantum systems to external environments destroys quantum effects, such as quantum superpositions and entanglement. The microscopic picture is that this coupling generates correlations between the system and the environmental degrees of freedom~\cite{Zurek}. This results in an increase of the system's entropy, while the system state usually reaches a stationary state that is often well approximated by a thermal state~\cite{Winter,goldstein2006}. Such dynamics is well exemplified by the quantum Brownian motion, which can be microscopically modeled by the coupling of an oscillator embedded in an ion crystal \cite{rubin1963,weiss1999,ford1965}. In Ref.~\cite{rubin1963}, Rubin derived the conditions under which a defect oscillator thermalizes with the rest of the chain, which has been initially prepared in a thermal state at temperature $T$. This model provides an interesting realization of an Ohmic reservoir that contains in a natural way the relevant frequency scales. The physical system is closed and composed by one defect and the chain. From this perspective it is important to mention most recent studies that analyze thermalization in closed systems \cite{Eisert02,Riera11,rigol2008}, as well as recent proposals for simulating Ohmic reservoirs with chains of oscillators~\cite{Chin}.

Scaling up these dynamics by increasing the number of defects embedded in the crystal can lead to some surprises. Let us first assume that the parameters are chosen such that a single defect thermalizes with the rest of the chain. Contrary to the naive expectation that the two defects will reach a thermal state independent of their initial state, the two defects can be entangled by the reservoir at sufficiently low temperatures, even if they have been initially prepared in a separable state. 
This result can be ascribed to symmetries of the total Hamiltonian that effectively decouple collective variables of the defect oscillators from the rest of the chain, leading to so-called decoherence free subspaces~\cite{Lidar}. This mechanism of entanglement generation between two objects that are not coupled directly, but indirectly via a common larger physical system, has been discussed in various settings, see for instance \cite{huelga2002,braun2002,benatti2003,plenio2004,benatti2006,venuti2006,
jun_hong_an2007,hoerhammer2008,paz2008,zell2009,paz2009,guinea2009,cormick2010,benatti2010}. An important characteristic of most of these theoretical studies is the assumption that the two objects couple to an idealized bath with artificially chosen spectral density. By contrast, in Ref.~\cite{wolf2011} we considered the model of a one-dimensional harmonic crystal, whose spectral density was determined from {\it ab initio} calculations, and we showed that entanglement between distant defects can be generated by the excitations of the crystal. However, a harmonic crystal does not always act as a perfect bosonic heat bath, since it can happen that the defect never relaxes to a steady state, even in the thermodynamic limit~\cite{rubin1963}. In the following, we perform a detailed investigation of the conditions under which (i) a generic harmonic chain plays the role of a thermalizing heat bath and (ii) two harmonic defects that couple to this chain are found to be entangled in the steady state. To this end, we numerically integrate the exact Heisenberg equations of motion of the total system, without making any weak-coupling or Markovian approximations. This allows us to explore the full parameter regime.
\begin{figure}[h] 
\includegraphics[width=0.8\columnwidth]{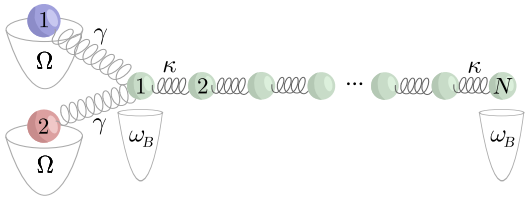}
\caption{(Color online) Sketch of the microscopic model used for entanglement generation between the two defect oscillators 1 and 2 (blue and red, respectively). The defects are confined by harmonic potentials with trap frequency~$\Omega$ and couple with the strength~$\gamma$ to a harmonic chain of $N$ particles. The particles of the chain interact via nearest-neighbor coupling with the strength~$\kappa$. The potentials with trap frequency~$\omega_B$ pin the edge oscillators of the harmonic chain.}
\label{fig1}
\end{figure}

In this work we extend and complement parts of the findings reported in Ref.~\cite{wolf2011} and systematically analyze the entanglement generation based on the microscopic model shown in Fig.~\ref{fig1}, where the two defects couple at the same site of the ion chain.  We examine the correlations between the defect oscillators for time scales that are smaller than the recurrence time (due to finite size effects), but for which a (quasi) stationary state is reached. Our objective is to connect our model predictions with previous studies on similar systems that were based on a phenomenological description of the reservoir \cite{huelga2002,benatti2006,hoerhammer2008,paz2008,paz2009,zell2009}. For this purpose we tune the parameters to a regime in which the chain effectively behaves like a (quasi) Ohmic reservoir. The numerical study allows us to determine both the stationary state, if it exists, as well as the out-of-equilibrium dynamics for a vast range of parameters, for which a master equation description of the defect dynamics may not be convenient. The simulations are supported by analytical investigations that yield a general criterion for the existence of steady-state entanglement.

This paper is organized as follows: The microscopic model at the basis of our analysis is introduced in Sec.~\ref{Sec:MicroModel}. Here, the basic idea leading to entanglement generation mediated by the chain is sketched. Section \ref{Sec:Form} describes the theoretical formalism. The dynamics of the defects is studied in Sec.~\ref{Sec:reservoirCharcterization} by means of a generalized quantum Langevin equation. The spectral density of the chain is discussed and the parameter regimes for which the harmonic chain acts as an Ohmic reservoir are identified. In Sec.~\ref{sec:LogarithmicNeg} a detailed analysis of the entanglement behavior for different initial conditions and coupling parameters is given. The conclusions are drawn in Sec.~\ref{Sec:Conclusions}, and the Appendixes~\ref{App:NewSqueezingParameters}--\ref{App:LogNeg} provide further aspects, as well as details of the calculations related to the Sec.~\ref{Sec:Form}--\ref{sec:LogarithmicNeg}. 


\section{Entanglement mediated by the chain\label{Sec:MicroModel}}

In this section we first introduce the microscopic model that provides the basis of our study on entanglement generation between two oscillators and then present the main idea why the two defect oscillators can become entangled via the interaction with the ion chain. 

The physical system is illustrated in Fig.~\ref{fig1}. It  is composed of a chain of $N+2$ oscillators that couple with nearest-neighbor interaction. Among these, $N$ oscillators have mass $m$ and form an ordered linear chain with interparticle distance $a$ and interparticle coupling strength~$\kappa$. The oscillators at both ends of the chain are pinned by harmonic traps with frequency~$\omega_B$.
The two additional defects have mass $M$ and are confined by a harmonic potential with trap frequency~$\Omega$. They couple with the same strength $\gamma$ to the oscillator at one edge of the chain. The chain has been prepared in a thermal state at temperature $T$. Our objective is to determine under which conditions the defect oscillators are entangled in the steady state.


\subsection{Hamiltonian\label{Sec:Hamiltonian}}

The Hamiltonian determining the dynamics of the chain and the defect oscillators reads
\begin{equation}
H = H_S+H_B+H_I
\label{eq:H:0}
\end{equation}
and comprises the free Hamiltonian of the two defect oscillators,
\begin{equation}
H_S=\sum_{\mu=1}^2 \left[ \frac{\PP^2_\mu}{2 M} +\frac 12 M\Omega^2 \XX^2_\mu \right]\,,
\label{eq:HSF}
\end{equation}
the free Hamiltonian of the reservoir,
\begin{equation}
H_B=\sum_{i=1}^{N}\left[\frac{p_i^2}{2 m}+\frac{m}{2}\omega_i^2\, x_i^2\right]
+\frac{\kappa}{2}\sum_{i=1}^{N-1}(x_i- x_{i+1})^2\,,
\label{eq:HBF}
\end{equation}
and the interaction Hamiltonian,
\begin{equation}
H_I=\frac\gamma 2 \Big[(X_1-x_1)^2+(X_2-x_1)^2\Big]\,,
\label{eq:HIF}
\end{equation}
which is assumed to be switched on at $t=0$.

Here, $\XX_\mu$ denotes the position of the defect \mbox{($\mu=1,2$)}, and $x_i$ the displacement of the chain particle from the equilibrium position $x_i^ {(0)}=ia$ (${i\in\{1,\ldots N\}})$. With the corresponding canonically conjugate momenta $\PP_\mu$ and $p_i$, the nonvanishing commutation relations read ${[\XX_\mu,\PP_\mu]={\rm i}\hbar}$ and $[x_i,p_i]={\rm i}\hbar$. Moreover, the shorthand notation ${\omega_i=\omega_B (\delta_{i,1}+\delta_{i,N})}$ incorporates the trap frequencies of the edge oscillators in the chain.

\subsection{Basic idea of entanglement generation}

In presence of only one defect oscillator, the model in Fig.~\ref{fig1} provides a generalization of Rubin's model~\cite{rubin1963}. Rubin showed in particular that the chain can act as a thermal bath for a single defect, provided some conditions are fulfilled, which involve the ratio $M/m$ between the defect and the ions masses, the strength of the coupling, and the time scales in which the dynamics are analyzed. The scope of Sec.~\ref{Sec:reservoirCharcterization} is to determine under which specific conditions this dynamics is encountered for a finite chain. In this section we focus on the general idea and show that the ion chain can create entanglement between two defects, which are initially prepared in an uncorrelated quantum state. 

In general, bath-induced entanglement is endorsed by the symmetries of the Hamiltonian or, in the case of open quantum systems, by the symmetries of the master equation. We first observe that the total Hamiltonian~\eqref{eq:H:0} is invariant under exchange of the coordinates of the two defect oscillators. It is therefore convenient to introduce center-of-mass (COM) and relative coordinates for the defect particles, $$X_\pm=(X_1\pm X_2)/\sqrt{2}\,,$$ 
and the corresponding canonically conjugate momenta, $P_\pm=(P_1\pm P_2)/\sqrt{2}$, where the subscript $+$ ($-$) denotes the COM (relative) motion. In this representation, the Hamiltonian~\eqref{eq:H:0} can be written as the sum ${H= H_-+H_+}$, where 
\begin{equation}
H_-=\frac{P^2_-}{2M} + \frac{1}{2}M\Omega_\gamma^ 2 X^2_-
\label{eq:Hminus}
\end{equation}
governs the dynamics of the relative motion, and
\begin{equation}
H_+=\frac{P^2_+}{2M} + \frac{1}{2}M\Omega_\gamma^ 2 X_+^2+H_B'-(\sqrt{2}\gamma)\, X_+\, x_1
\label{eq:Hplus}
\end{equation}
describes the coupling of the COM motion to the chain. Here, we denoted by
\begin{equation}
\Omega_\gamma=\sqrt{\Omega^2+\gamma/M}
\label{Omega:gamma}
\end{equation}
the shifted trap frequency and by 
$$H_B'=H_B+\gamma x_1^2$$
the chain Hamiltonian that includes the effect of the coupling constant $\gamma$ on the eigenspectrum. In this form it is evident that $H_-$ is a constant of motion: The relative motion is decoupled from the chain. The COM, on the other hand, behaves as an effective defect particle that couples to one edge of the chain with the coupling strength $\sqrt{2}\gamma$. 

Under the conditions for which the chain acts as thermal bath for a single defect, it will induce thermalization of the COM defect particle and wash out possible initial correlations between COM and relative motion of the defects. While the COM approaches a thermal state at temperature $T$ after a transient time, the relative motion evolves freely and preserves some features of the initial states of the defects.

\begin{figure}[h] 
\includegraphics[width=\columnwidth]{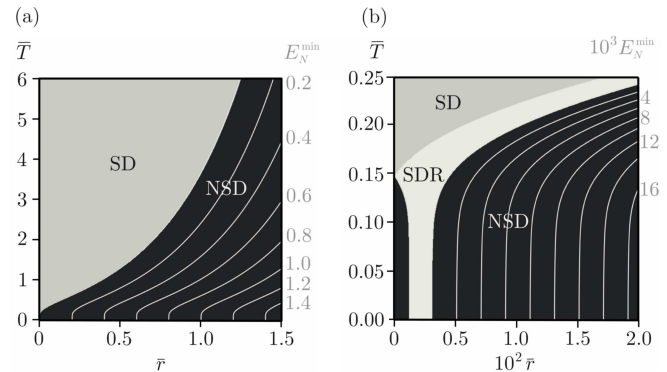}
\caption{Contour plots of the logarithmic negativity~$E_N(\di r,\di T)$ as a function of the initial squeezing parameter $\bar r$ of the oscillators and of the temperature $\di T=T/T_S$, with $T_S=\hbar\Omega_\gamma/k_B$ defining the temperature scale. The contour plot is evaluated for a chain of 1000 ions at times at which the COM motion has reached a stationary state. (b) Behavior at low temperatures. The black regions that are denoted by "NSD" indicate the parameter regime in which the defect oscillators are entangled in their steady state. The parameters are $m=0.5M$, $\gamma=0.1\kappa$, and $\kappa=M\Omega_\gamma^2$. The squeezing parameters of both oscillators are taken to be equal. Further details are provided in Sec. \ref{sec:LogarithmicNeg}.}
\label{fig9} 
\end{figure}

The above-described dynamics is the key point in the creation of steady-state entanglement between the defects. For instance, if the relative motion is in a squeezed state and the temperature of the COM is sufficiently low, the product of the two orthogonal quadratures $\Delta X_-\Delta P_+$ (here taken in the reference frame rotating at the oscillator frequency $\Omega_\gamma$) can fall below the standard quantum limit giving rise to two-mode squeezing of the defects~\cite{Reid2009} and thus entanglement. The squeezing of the relative coordinate can be easily realized by preparing each individual defect in a squeezed state at the time $t=0$. Figure~\ref{fig9} displays the contour plot of the logarithmic negativity~\cite{vidal2002,plenio2005} that quantifies the entanglement between the defect oscillators. The logarithmic negativity is shown as a function of the chain temperature $T$ and of the initial squeezing parameter $r$ of each defect oscillator~\cite{paz2008,paz2009}. The details of the calculations are provided in Sec.~\ref{sec:LogarithmicNeg}.

We note that this kind of dynamics has been predicted in Refs.~\cite{paz2008,paz2009}, where contour plots like the one in Fig.~\ref{fig9} have been introduced for the first time. In contrast to our work, the model used in Refs.~\cite{paz2008,paz2009} takes advantage of the Hu-Paz-Zhang master equation~\cite{hu1992} and is based on a phenomenological treatment of the bath. In the present work, the bath is modeled by a chain of harmonic oscillators. Although we investigate a parameter regime in which our microscopic system reproduces the results of the Hu-Paz-Zhang master equation, we could likewise consider entanglement generation for a parameter regime in which the COM motion does not reach a thermal state. Such a regime, however, lies beyond the description based on the Hu-Paz-Zhang master equation~\cite{hu1992,ford2001}.

We also would like to mention that entanglement mediated by a chain of oscillators has been investigated in a series of works, such as \cite{audenaert2002,plenio2004,anders2008A,anders2008B,plenio2005B}. In these works the chain is a homogeneous one-dimensional crystal and thus possesses discrete translational invariance. The regime is such that a unique stationary state exists in the thermodynamic limit which corresponds to a thermal state \cite{ford1965}.  In Refs. \cite{audenaert2002,anders2008A,anders2008B} the authors characterize entanglement between two components of the chain in the steady state. The entanglement found in \cite{plenio2004,plenio2005B} between the ions at the chain edges is instead a dynamical effect, which obviously vanishes in the thermodynamic limit.

\section{Theoretical formalism\label{Sec:Form}}

In this section we develop the mathematical formalism, which we employ in the following sections for the characterization of the chain and for the analysis of the steady-state entanglement between the defects.

For later convenience, we introduce the vector operators for the reservoir particles ${\xx^\tp=(x_{1},\ldots,x_N)}$ and ${\pp^\tp=(p_{1},\ldots,p_N)}$ and rewrite
the reservoir Hamiltonian~\eqref{eq:HBF} in the form
\begin{equation}
H_B=\frac{\pp^2}{2m}+\frac12\,\xx^\tp V\xx
\label{eq:HBFmatrix}
\end{equation}
with the potential matrix $V\in\mathds{R}^{N\times N}$ given by
\begin{equation}
 V=\left(
\begin{array}{ccccc}
m\omega_B^2+\kappa & -\kappa &  &  &   \\
\phantom{m\omega_B^2 }-\kappa & 2 \kappa & -\kappa &  &   \\
  & \ddots & \ddots & \ddots &   \\
  &  & -\kappa & 2\kappa & -\kappa\phantom{m\omega_B^2} \\
  &  &  &  -\kappa & m\omega_B^2+\kappa
\end{array}
\right)\,.
\label{eq:V}
\end{equation}
The coupling between the oscillators and the reservoir induces a shift in the trap frequencies of the defect and chain particles, that depends on the coupling strength $\gamma$. This effect can be highlighted by rewriting the full Hamiltonian~\eqref{eq:H:0} in the form
\begin{equation}
\begin{split}
H=&\sum_{\mu=1}^2 \left[\frac{P^2_\mu}{2 M} + \frac{M}{2}\Omega_\gamma^2X^2_\mu \right]
+\frac{\pp^2}{2 m}+\frac{1}{2}\,\xx^\tp\, V^{\basind{\gamma}}\,\xx\\
&-\gamma\,x_1 \left(X_1+ X_2\right)\,,
\end{split}
\label{eq:Hdim:0}
\end{equation}
where 
\begin{equation}
 V^{\basind{\gamma}}=V+{2\gamma}\,\ee_1\otimes \ee_1^\tp
\label{eq:Vgamma}
\end{equation}
denotes the potential matrix including the shift due to the interaction. The quantity ${\ee^\tp_1=(1,0,\ldots,0)\in\mathds{R}^N}$ is the first unit vector and $\otimes$ represents the dyadic product. 

An important point consists of the boundary conditions. For the model under consideration, we assume that the oscillators at both ends of the chain are confined by harmonic potentials with frequency $\omega_B$. Although the potential of the ion at the other chain edge, $j=N$, has no influence on the dynamics of the defects for the time scales which are relevant to our analysis, we include it for symmetry reasons. As long as not specified elsewhere, we assume that ${\omega_B=\sqrt{\kappa/m}}$ throughout this paper. 

\subsection{Initial states\label{Sec:Initial}}

The initial state of the defect oscillators and the chain is given by the density matrix
\begin{equation*}
\chi(0)=\rho_1\otimes\rho_2\otimes\rho_B(T)\,,
\end{equation*}
where $\rho_{\mu}$ denotes the state of the defect oscillator (${\mu=1,2}$) and 
\begin{equation}
\rho_B(T)=\exp\left(-\beta H_B\right)/Z
\label{eq:InitialStateChain}
\end{equation}
describes the thermal state of the chain at temperature~$T$. Here, $Z={\rm Tr}\{\exp\left(-\beta H_B\right)\}$ is the partition function and $\beta= (k_BT)^ {-1}$ the inverse temperature with $k_B$ as Boltzmann constant. Due to this choice of $\chi(0)$, there exist neither correlations between the defect oscillators nor between the defects and chain at $t=0$.

More specifically, the defect oscillators are assumed to be prepared in pure states ${\rho_\mu=|s_\mu\rangle\langle s_\mu|}$. Here,~$|s_\mu\rangle$ denotes a squeezed state whose squeezing parameter ${s_\mu=r_\mu\,{\rm e}^{{\rm i}\phi_\mu}}$ is given by the absolute value ${r_\mu\geq 0}$ and the angle ${\phi_\mu\in(-\pi,\pi]}$. The corresponding first and second moments read $\ave{X_\mu}=\ave{P_\mu}=0$ and 
\begin{align}
\left[\sigma_\mu(0)\right]_{11}& =\ave{X_\mu^2}-\ave{X_\mu}^2\nonumber\\
             &=\frac{\hbar}{2M \Omega}\left({\rm e}^{-2 r_\mu}\cos^2\frac{\phi_\mu}{2}
+{\rm e}^{2 r_\mu}\sin^2\frac{\phi_\mu}{2}\right)\,,
\label{eq:sigma11}\\
\left[\sigma_\mu(0)\right]_{22}& =\ave{P_\mu^2}-\ave{P_\mu}^2\nonumber\\
             &=\frac{\hbar M \Omega}{2}\left({\rm e}^{-2 r_\mu}\sin^2\frac{\phi_\mu}{2}+{\rm e}^{2 r_\mu}\cos^2\frac{\phi_\mu}{2}\right)\,,
\label{eq:sigma22}\\
\left[\sigma_\mu(0)\right]_{12}& =\frac12\ave{X_\mu P_\mu+P_\mu X_\mu}-\ave{X_\mu}\ave{P_\mu}
\nonumber\\
             &=-\frac{\hbar}{2}\sinh(2 r_\mu)\sin\phi_\mu\,,
\label{eq:sigma12}
\end{align}
with $\ave{\cdot}={\rm Tr}\{\cdot \chi(0)\}$. The moments in Eqs. \eqref{eq:sigma11}-\eqref{eq:sigma12} define the initial covariance matrices $\sigma_\mu(0)$ of the defect oscillators at the time $t=0$. 

For the following analysis it is also convenient to introduce the initial covariance matrix of the harmonic chain. We express it in terms of the individual block matrix elements
\begin{align*}
\sigma_{\xx\xx}(0)&=\ave{\xx\otimes\xx^\tp}-\ave{\xx}\otimes\ave{\xx}^\tp\,,\\
\sigma_{\pp\pp}(0)&=\ave{\pp\otimes\pp^\tp}-\ave{\pp}\otimes\ave{\pp}^\tp\,,\\
\sigma_{\xx\pp}(0)&=\frac12\ave{\xx\otimes\pp^\tp+\pp\otimes\xx^\tp}-\ave{\xx}\otimes\ave{\pp}^\tp\,,
\end{align*}
whose explicit forms depend on the potential matrix~\eqref{eq:V} and read~\cite{ford1965}
\begin{align}
 \sigma_{\xx\xx}(0)&=\frac{\hbar}{2}(m V)^{-\frac{1}{2}}\,
\coth\bigg(\frac{\beta\hbar}{2}\bigg(\frac{V}{m}\bigg)^{\!\frac12}\,\bigg)\,,
\label{eq:initialCovarianceBath}\\
 \sigma_{\pp\pp}(0)&=\frac{\hbar}{2}(m V)^{\frac{1}{2}}\,
\coth\bigg(\frac{\beta\hbar}{2}\bigg(\frac{V}{m}\bigg)^{\!\frac12}\,\bigg)\,,
\nonumber
\end{align}
together with ${\sigma_{\xx\pp}(0)=0}$ and ${\ave{\xx}=\ave{\pp}=0}$. 


\subsection{Dimensionless variables\label{Sec:DimensionlessDes}}

With the total Hamiltonian and the initial covariance matrices at hand, we now introduce a dimensionless description of our microscopic model. This reformulation is useful for the determination of the logarithmic negativity between the two defects.

A typical length scale is the width of the ground state of the defect oscillator Hamiltonian~\eqref{eq:HSF}, $${\alpha_\gamma=\sqrt{\hbar/(M\Omega_\gamma)}}\,.$$ 
The dimensionless position and momentum operators for the two defects are defined as ${\di X_\mu=X_\mu/\alpha_\gamma}$, ${\di P_\mu=\alpha_\gamma P_\mu/\hbar}$. For the oscillators of the reservoir we accordingly define ${\di x_i=x_i/\alpha_\gamma}$ and ${\di p_i=\alpha_\gamma p_i/\hbar}$. These definitions imply the nonvanishing commutation relations
$$[\di X_\mu,\di P_\mu]={\rm i}=[\di x_i,\di p_i]\,.$$ 

We further introduce the dimensionless mass $\di m$, trap frequencies $\di \omega_B$ and $\di\Omega$, and coupling constants~${\di\kappa}$ and ${\di\gamma}$ according to
\begin{eqnarray*}
&&\di m=m/M\,,\\
&&\di\omega_B=\omega_B/\Omega_\gamma\,,~~~~\di\Omega=\Omega/\Omega_\gamma\,,\\
&&\di\kappa=\kappa/(M\Omega^2_\gamma)\,,~~~\di\gamma=\gamma/(M\Omega^2_\gamma)\,.
\end{eqnarray*}
With this choice, the mass of the defects $M$ defines the unit mass, the shifted frequency $\Omega_\gamma$, Eq.~\eqref{Omega:gamma}, is the unit frequency, and the energy $M\Omega_\gamma^2$ sets the relevant energy scale. We note that the rescaled coupling strength ${\di\gamma=\gamma/(\gamma+M\Omega^2)}$ can only take on values in the interval
$0\le \di\gamma<1$. Here, $\di\gamma=0$ corresponds to $\gamma=0$, while $\di\gamma\to 1$ represents the limit of infinitely large coupling~${\gamma\to\infty}$.

The rescaled Hamiltonian ${\di H=H/(\hbar\Omega_\gamma)}$ then reads
\begin{equation}
\begin{split}
\di H=&\frac12\sum_{\mu=1}^2 \left[\di P^2_\mu + \di X^2_\mu \right]
+\frac{\di \pp^2}{2\di m}+\frac12\,\di\xx^\tp\,\di V^{\basind{\gamma}}\,\di\xx\\
&-\di\gamma\,\di x_1 \left(\di X_1+ \di X_2\right)\,,
\end{split}
\label{eq:Hdim}
\end{equation}
with ${\di V^{\basind{\gamma}}=V^{\basind{\gamma}}/(M\Omega^2_\gamma)}$. The rescaled time is given by the variable $${\di t=\Omega_\gamma\,t}\,.$$ 

For later convenience we also report the Hamiltonians governing the dynamics of relative and COM motion in their dimensionless form. They are given by
\begin{equation}
\di H_-=\frac12 \left(\di P^2_- + \di X^2_-\right)
\label{eq:Hminus:di}
\end{equation}
and
\begin{equation}
\di H_+=\frac12 \left(\di P^2_+ + \di X^2_+\right)
+\frac{\di \pp^2}{2\di m}+\frac12\,\di\xx^\tp\,\di V^{\basind{\gamma}}\,\di\xx
-\di X_+\left(\di\bgamma^\tp\di\xx\right)\,,
\label{eq:Hplus:di}
\end{equation}
where we have introduced the dimensionless coupling vector ${\di\bgamma^\tp=(\sqrt{2}\,\di \gamma,0,\ldots,0)\in\mathds{R}^{N}}$. 

According to these definitions, an operator function $f(X_\mu,P_\mu;x_i,p_i)$ acting on the Hilbert space of the total system takes the rescaled form ${\di f=f(\di X_\mu,\di P_\mu;\di x_i,\di p_i)}$ and satisfies the Heisenberg equation 
\begin{equation*}
 \diff{\di f}{\di t}={\rm i} \,[\di H, \di f]\,.
\end{equation*}

We now come to the rescaled covariance matrices. With the dimensionless temperature ${\di T= k_B T/(\hbar\, \Omega_\gamma)}$, the inverse temperature ${\di \beta=\di T^{-1}=\beta\,\hbar\, \Omega_\gamma}$, and the potential matrix ${\di V=V/(M\Omega^2_\gamma)}$, the nonvanishing block matrix elements~\eqref{eq:initialCovarianceBath} read in dimensionless form
\begin{align}
\di\sigma_{\dxx\dxx}&=\frac{1}{2}(\di m \di V)^{-\frac{1}{2}}\,
\coth\bigg(\frac{\di \beta}{2}\bigg(\frac{\di V}{\di m}\bigg)^{\!\frac12}\bigg)\,,
\label{eq:CovarianceBathdim}\\
\di \sigma_{\dpp\dpp}&=\frac{1}{2}(\di m \di V)^{\frac{1}{2}}\,
\coth\bigg(\frac{\di \beta}{2}\bigg(\frac{\di V}{\di m}\bigg)^{\!\frac12}\bigg)\,.
\nonumber
\end{align}
Based on an appropriate one-to-one mapping $r=r(\di r,\di \phi)$ and ${\phi=\phi(\di r,\di \phi)}$ between the original and the new squeezing parameters ${\di r_\mu\geq 0}$, ${\di \phi_\mu\in (-\pi,\pi]}$, the covariance matrices for the defect oscillators~\eqref{eq:sigma11}-\eqref{eq:sigma12} can be expressed in the convenient form 
\begin{equation}
\di\sigma_\mu(0)=\frac12 \,O^\tp(\di \phi_\mu)
\,S(e^{2 \di r_\mu})\,O(\di\phi_\mu)\,.
\label{eq:CovarianceSystemdimStandard}
\end{equation}
In this expression, we introduced the symplectic and orthogonal matrices ($z\in\mathds{C}$)
\begin{equation}
S(z)=\left(
\begin{array}{cc} 
z^{-1} & 0 \\
0 & z
\end{array}
\right)\;\;\text{and}\;\;
O(\phi)=\left(
\begin{array}{cc} 
\phantom{-}\cos\frac\phi2 & \sin\frac\phi2 \\
-\sin\frac\phi2 & \cos\frac\phi2
\end{array}
\right)\,.
\label{eq:DefOandS}
\end{equation}
In this way, the elements of the initial covariance matrix for the defect oscillators~\eqref{eq:sigma11}-\eqref{eq:sigma12} reduce to
\begin{align*}
\left[\di\sigma_\mu(0)\right]_{11}& =\frac{1}{2}\left({\rm e}^{-2 \di r_\mu}\cos^2\frac{\di \phi_\mu}{2}
+{\rm e}^{2 \di r_\mu}\sin^2\frac{\di\phi_\mu}{2}\right)\,,\\
\left[\di\sigma_\mu(0)\right]_{22}& =\frac{1}{2}\left({\rm e}^{-2 \di r_\mu}\sin^2\frac{\di\phi_\mu}{2}+{\rm e}^{2 \di r_\mu}\cos^2\frac{\di \phi_\mu}{2}\right)\,,\\
\left[\di\sigma_\mu(0)\right]_{12}& =-\frac{1}{2}\sinh(2 \di r_\mu)\sin\di\phi_\mu\,.
\end{align*} 
The above-mentioned one-to-one mapping is discussed in detail in Appendix~\ref{App:NewSqueezingParameters}. The new parameters $\di r_\mu$ and $\di \phi_\mu$ define the squeezing of the defect oscillators with respect to the shifted trap frequency~$\Omega_\gamma$. Therefore, the squeezing parameter $\di r=0$ corresponds to the ground state of a harmonic oscillator with trap frequency~$\Omega_\gamma$.


\subsection{Formal solution of the equations of motion\label{Sec:FormalSolutionHeisenberg}}

The formal solution of the Heisenberg equations of motion for the position and momentum operators of both defect and bath oscillators can be written as a linear map between their initial and final values. For this purpose, we introduce the vector of the position and momentum operators for defect and chain oscillators,  ${\bzeta^\tp=(\di X_1,\di P_1,\di X_2,\di P_2,\dxx^\tp,\dpp^\tp)\in \mathds{R}^{4+2N}}$, and rewrite the total Hamiltonian~\eqref{eq:Hdim} as ${\di H=\frac12\,\bzeta^\tp\di{\mathcal{H}}\,\bzeta}$, with the positive definite matrix~$\di{\mathcal{H}}$. Furthermore, we introduce the antisymmetric block matrix
\begin{equation*}
 \mathcal{J}=\left(
\begin{array}{cc}
J_2 & 0 \\
0 & J_N
\end{array}
\right)=-\mathcal{J}^\tp=-\mathcal{J}^{-1}
\end{equation*}
that contains the submatrices
\begin{equation}
J_2=\left(
\begin{array}{cccc}
\phantom{-}0 & 1 & 0 & 0 \\
-1 & 0 & 0 & 0 \\
0 & 0 & \phantom{-}0 & 1 \\
0 & 0 & -1 & 0 
\end{array}\right)\quad\text{and}\quad
J_N=\left(
\begin{array}{cc}
\phantom{-}0 & \mathds{1}_N\\
-\mathds{1}_N & 0
\end{array}\right)\,.
\label{eq:DefJ}
\end{equation}
Here, $\mathds{1}_N\in \mathds{R}^{N\times N}$ denotes the identity matrix.

By means of these definitions, the Heisenberg equations of motion for the position and momentum operators reduce to
\begin{equation*}
\diff{\bzeta}{\di t}=i \,[\di H, \bzeta(\di t)] = \mathcal{J}\,\di{\mathcal{H}}\,\bzeta(\di t)\,.
\end{equation*}
Their formal solution reads
\begin{equation}
\bzeta(\di t)=\mathcal{T}(\di t)\,\bzeta(0)\,,
\label{eq:FormalSolutionHeisenberg}
\end{equation}
with the symplectic matrix
\begin{equation*}
\mathcal{T}(\di t)=e^{\mathcal{J}\di{\mathcal{H}}\,\di t}\,.
\end{equation*}
The time evolution of the total covariance matrix, $\totalVar(\di t)$, is given in terms of the linear mapping by the relation 
\begin{equation}
 \label{V:tot}
 \totalVar(\di t)=\mathcal{T}(\di t)\,\totalVar(0)\,\mathcal{T}^\tp(\di t)\,,
\end{equation}
where $\totalVar(0)$ is the total covariance matrix at $t=0$, which is composed of the initial covariance matrices~\eqref{eq:CovarianceBathdim} and~\eqref{eq:CovarianceSystemdimStandard} and takes the form
\begin{equation*}
\totalVar(0)=\left(
\begin{array}{cccc}
 \di \sigma_1(0) & 0 & 0 & 0 \\
0 & \di \sigma_2(0) & 0 & 0 \\
0 & 0 & \di \sigma_{\dxx\dxx}(0) & 0 \\
0 & 0 & 0 & \di \sigma_{\dpp\dpp}(0)
\end{array}
\right)\,.
\end{equation*}
Equation~\eqref{V:tot} represents the basis of the numerical simulations used in the analysis of entanglement generation. In this context, the covariance matrix of the defect oscillators $\di\Sigma (\di t)$ is of particular interest. It is extracted from the total covariance matrix~$\totalVar(\di t)$ according to
\begin{equation}
\label{eq:sigma:t}
\left[\di \Sigma (\di t)\right]_{ik}=\left[\totalVar (\di t)\right]_{ik}\,,
\end{equation}
with $i,k\in\{1,2,3,4\}$. Since we aim at the determination of the steady-state entanglement, it suffices to evaluate the covariance matrix~$\di\Sigma (\di t)$ at times $\di t>\tth$. Here, $\tth$ represents the time scale at which the COM defect oscillator reaches a stationary state, provided the harmonic chain acts as a thermal bath. For this reason, we examine in the next section the conditions for which the reservoir displays this behavior.


\section{Characterization of the reservoir \label{Sec:reservoirCharcterization}}

The harmonic chain plays a basic role in our study of entanglement generation between the defects for the following reason: Although the total dynamics is unitary and the system is finite, the chain can act as a thermal bath for the COM motion of the defects, while the relative motion is uncoupled. 
In order to understand under which conditions this mechanism leads to entanglement, a detailed knowledge about the action of the chain on the COM motion is necessary. Hence, the purpose of this section is to characterize the chain in terms of a reservoir and identify the parameter regime for which these conditions are valid.

\subsection{Generalized Quantum Langevin Equations for the defects}

Let us consider the dynamics of the defect oscillators. The dynamics of the relative motion is governed by the Hamiltonian \eqref{eq:Hminus:di}, and the solution of the corresponding Heisenberg equations of motion simply describes the evolution of a harmonic oscillator with frequency $\Omega_\gamma$, that reads
\begin{align}
\di X_-(\di t)&=\di X_-(0) \cos\di t+\di P_-(0) \sin\di t\,,
\label{eq:relativeMotionModel1}\\
\di P_-(\di t)&=-\di X_-(0) \sin\di t+\di P_-(0) \cos\di t\,,
\nonumber
\end{align}
where we recall that $\di t=\Omega_\gamma t$. The COM motion, nevertheless, remains coupled to the oscillator at the chain edge. We rewrite its equation of motion in terms of a generalized quantum Langevin equation (GQLE). Starting from the Heisenberg equations of motion for the operators $\di X_+$, $\di P_+$, $\di\xx$, and~$\di \pp$, the GQLE follows by formal integration of the equations for the chain degrees of freedom~\cite{weiss1999} and takes the form
\begin{align}
\diffz{\di X_+}{\di t}& + \!\int\limits_0^{\di t}\di\Gamma_+(\di t-t')\,\diff{\di X_+}{ t'}\,d t'
+(1-\di\Gamma_+(0))\,\di X_+(\di t)
\nonumber\\
&=\di F_+(\di t)-\di\Gamma_+(\di t)\,\di X_+(0)\,.
\label{eq:GQLEModel1}
\end{align}
Here, we have introduced the memory-friction kernel~\cite{weiss1999}, which reads 
\begin{equation}
\di\Gamma_+(\di t)=
\sum_{j=1}^{N}\frac{(\di\gamma^{+}_j)^2}{\di m (\di\omega^{+}_j)^2}\cos(\di\omega^+_j \di t)
\label{eq:frictionKernelModel1}
\end{equation}
for $\di t\geq 0$, while it vanishes otherwise. We have also introduced the operator-valued random force, which is defined by~\cite{weiss1999}
\begin{equation}
\di F_+(\di t)=\!
\sum_{j=1}^{N}\!\left[\di\gamma^{+}_j \cos(\di\omega^{+}_j \di t)\, \di x^{+}_j(0)+
\frac{\di\gamma^{+}_j}{\di m \di \omega^{+}_j} \sin(\di\omega^{+}_j  \di t)\, \di p^{+}_j(0)\right]\!.
\label{eq:randomForceModel1}
\end{equation}
In the expressions for the memory-friction kernel and the random force, the quantities~$\di\omega^+_j$ and~$\di \gamma^+_j$ appear. The first ones denote the eigenfrequencies of the potential matrix $\di V^{\basind{\gamma}}$ given by Eq.~\eqref{eq:Vgamma}. They follow from the diagonalization of the chain potential $V^{\basind{\gamma}}$ and are defined by the relation 
\begin{equation}
O_+^\tp \,\di V^{\basind{\gamma}} O_+=\di m\cdot{\rm diag}((\di\omega_1^+)^2,\ldots,(\di\omega^+_{N})^2)\,,
\label{spectrum}
\end{equation}
where $O_+$ is the orthogonal matrix which diagonalizes $\di V^{\basind{\gamma}}$.  In particular, the  orthogonal matrix~$O_+$  establishes the relation between the normal and the original coordinates, ${\di \xx_+=O_+^\tp\,\di\xx}$ and ${\di \pp_+=O_+^\tp\,\di\pp}$, see e.g.~\cite{Goldstein}. The quantities $\di x^{+}_j(0)$ and $\di p^{+}_j(0)$ in Eq.~\eqref{eq:randomForceModel1} stand for the $j$-th component of the vectors $\di \xx_+$ and $\di \pp_+$, respectively. The parameters $\di \gamma^+_j$ are the coupling strengths to the $j$-th normal mode of the reservoir and are given by ${\di\bgamma_+=O_+^\tp\,\di\bgamma}$. In the following we adopt the convention that the eigenfrequencies are ordered according to ${0<\di \omega_1^+<\di \omega_2^+<\ldots<\di \omega_{N}^+}$.

An important quantity that characterizes the influence of the reservoir on the COM motion is the environmental spectral density. This quantity is the Fourier cosine-transform of the memory-friction kernel~\eqref{eq:frictionKernelModel1}
\begin{eqnarray}
 \di J_{+}(\di\omega)
&=&\di\omega\,\int_0^\infty \di\Gamma_+(\di t)\cos(\di\omega \di t)\, {\rm d}\di t\,
\nonumber\\
&=& \frac{\pi}{2 } \sum_{j=1}^{N} 
\frac{(\di\gamma_j^+)^2}{\di m \,\di\omega^+_j}\,\delta(\di\omega-\di\omega^+_j)\,.
\label{eq:SpectralDensity}
\end{eqnarray}
The spectral density provides important insight into the action of the chain on the dynamics of the COM motion of the defect.

\begin{figure}[h]
\includegraphics[width=\columnwidth]{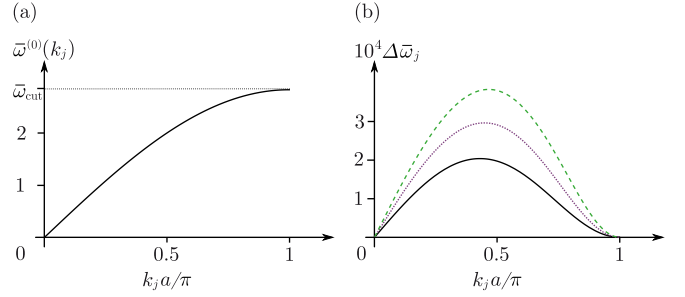}
\caption{(Color online) (a) Spectrum $\di\omega^{\basind{0}}_j$ of the potential $\di V$, Eq.~\eqref{eq:V}, for the parameter values $\di m=0.5$, $\di \gamma=0.1$, and $\di \kappa=1$ and (b) corrections $\Delta\di \omega_j$ to the eigenfrequencies of $\di V^{\basind{\gamma}}$ for the coupling strengths $\di \gamma=0.1$ (solid), $\di \gamma= 0.15$ (dotted), and $\di \gamma=0.2$ (dashed line). The chain is composed of $N=1000$ ions.}
\label{fig2}
\end{figure}

Before we proceed, we characterize the chain's normal modes. The eigenfrequencies are the solutions of Eq.~(\ref{spectrum}), which includes the shift due to the coupling of the defects with the edge ion. By appropriately setting the frequency of the edge potentials to the value $\di\omega_B=\sqrt{\di \kappa/\di m}$ (see Appendix~\ref{App:SpectralDensitySpecialCase}), we obtain for the normal mode spectrum in the limit $\di\gamma\to 0$
\begin{eqnarray}
\label{spectrum:0}
\di\omega^{\basind{0}}_j= \omcut\,\sin\left(\frac{ k_j a}2\right)\,,
\end{eqnarray}
where $\di\omega^{\basind{0}}_j\equiv \di\omega^{\basind{0}}(k_j)$ and ${k_j=j\pi/(a(N+1))}$ is the wave number, which appropriately denotes the modes when the Bloch theorem applies and takes on the values (${j\in\{1,\ldots ,N\}}$). This expression agrees with the one found for periodic boundary conditions~\cite{rubin1963}. The frequency ${\omcut=\sqrt{4\di\kappa/\di m}}$ is the high-frequency cutoff. The resulting spectrum, Eq. \eqref{spectrum:0}, is displayed in Fig.~\ref{fig2}. The eigenmodes, however, are, strictly speaking, not phononic waves.

Let us now consider the case $\di \gamma>0$. We expect for sufficiently small $\di\gamma$ that the effect of this coupling on the chain normal-mode spectrum is negligible. To quantify this statement, we consider the difference ${\Delta\di \omega_j=\di\omega^ +_j-\di\omega^{\basind{0}}_j}$ that involves the eigenfrequency $\di\omega^+_j$ given by Eq.~\eqref{spectrum} and the corresponding frequency $\di\omega^{\basind{0}}_j$ obtained in the limit $\di\gamma\to 0$. Figure~\ref{fig2}(b) displays the corrections $\Delta\di \omega_j$ for different coupling strengths $\di \gamma=0.1,0.15,0.2$ and constant $\di\kappa=1$. For these values, the difference $\Delta\di \omega_j$ is much smaller than all other physical parameters.

\begin{figure}[h]
\includegraphics[width=\columnwidth]{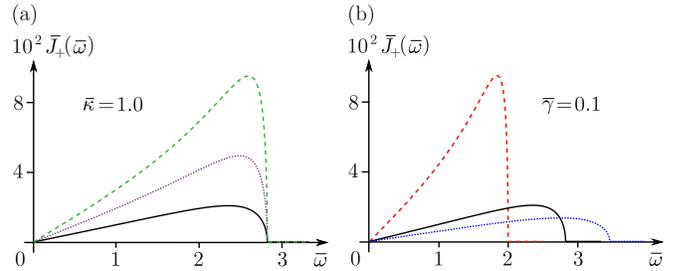}
\caption{(Color online) Spectral density $\di J_+(\di\omega)$ as a function of $\di\omega$. The parameters in (a) are $\di\kappa=1$, while $\di\gamma=0.1$ (solid), $\di\gamma=0.15$ (dotted), and $\di\gamma=0.2$ (dashed line). In (b) we take $\di \gamma=0.1$ while $\di\kappa=0.5$ (dashed), $\di\kappa=1$ (solid), and $\di\kappa=1.5$ (dotted). In both cases the mass ratio is $\di m=0.5$. The spectral densities is linear in the vicinity of the oscillator frequency~${\di\omega=1}$. A slight difference is found for the cases $\di \gamma=0.2$, $\di \kappa=1$ (a) and $\di \gamma=0.1$, $\di\kappa=0.5$ (b)  where $\di J_{+}(\di\omega)\sim \di \omega^a$ with $a>1$ in the vicinity of $\di \omega=1$. 
}
\label{fig3}
\end{figure}

Figures~\ref{fig3}(a) and~\ref{fig3}(b) display the spectral density for a choice of the parameters~$\di \gamma$ and $\di \kappa$ and taking $\di m=0.5$. For most of the parameter values the spectral density is linear about the value $\di\omega=1$, which corresponds to the frequency of the defect oscillator. In this case, the chain acts like a (quasi) Ohmic environment. A change in the mass ratio $\di m$ affects the spectral density in so far as the eigenfrequencies scales with $\di\omega^+_j\sim 1/\sqrt{\di m}$, leading to a change in the bandwidth $\bar \omega_{\rm cut}$.

\subsection{Thermodynamic limit}
\label{Sec:thermodynamic}

In Ref.~\cite{rubin1963} Rubin showed that a chain of oscillators with one embedded defect, exhibiting a spectrum as in Eq.~\eqref{spectrum:0}, can act like an Ohmic bath for the defect particle. This behavior is found in Rubin's model provided that the temperature of the chain is finite, the mass ratio satisfies $m/M<1$, and the thermodynamic limit is taken, which corresponds to the limit $N\to\infty$ keeping the interparticle distance $a$ in the chain constant~\cite{Footnote:Rubin}. Finite size effects are found for times larger than the time scale~$\trev$, which is discussed in the next section. However, they can be neglected if (i) the defect oscillator reaches a stationary state over time scales $\di t_{\rm th}$ such that $\di t_{\rm th}\ll \trev$ and (ii) the analysis can be restricted to these time scales.

We now discuss these assumptions in relation to our microscopic model, where, different from Rubin's model, the coupling strength $\gamma$ appears in addition to the coupling constant~$\kappa$. We are specifically interested in identifying the parameter regimes for which the effective defect of our model, the COM, thermalizes with the rest of the chain. 

For this purpose we first consider the formal solution of the Heisenberg equations of motion in Eq.~\eqref{eq:FormalSolutionHeisenberg}  in terms of the linear mapping~${\mathcal{T}_+(\di t)=e^{\mathcal{J}_+\di{\mathcal{H}}_+\,\di t}}$ (here the positive definite matrix $\di{\mathcal{H}}_+$ and the antisymmetric matrix $\mathcal{J}_+$ are defined in analogy to the discussion of Sec.~\ref{Sec:FormalSolutionHeisenberg}). The GQLE can formally be solved by applying a Laplace transformation to both sides of Eq.~\eqref{eq:GQLEModel1}, which yields an algebraic equation for the Laplace transform of~$\di X_+(\di t)$. In this case the residue theorem can be applied~\cite{footnote:residue}: The simple poles of the integrand are determined from the eigenfrequencies of the positive definite matrix ${\di W_+=\di T_+^{1/2}\,\di V_+\,\di T_+^{1/2}}$ with the block matrices
\begin{equation}
\di T_+=\left(
\begin{array}{cc}
1 & 0 \\
0 & \frac{\mathds{1}_N}{\di m}
\end{array}
\right)
\quad\text{and}\quad
\di V_+=\left(
\begin{array}{cc}
\phantom{-}1 & -\di\bgamma^\tp \\
-\di\bgamma & \phantom{-}\di V^{\basind{\gamma}}
\end{array}
\right)\,,
\label{eq:PotMatrixVPlus} 
\end{equation}
which respectively originate from the kinetic and potential energy part of the Hamiltonian $\di H_+$ in Eq.~\eqref{eq:Hplus}.
The sum of the residues yields a quasiperiodic function which is equivalent to the expression for~$\di X_+(\di t)$ deduced from the linear mapping $\mathcal{T}_+(\di t)$. As in Rubin's model of a single defect in a one-dimensional crystal~\cite{rubin1963}, thermalization is found in the limit $N\rightarrow \infty$ due to the formation of a continuous frequency band, provided no isolated frequencies above the frequency band occur. The existence of such isolated frequencies would result in residual oscillations of the coupled defect (in our case the COM motion) at long times. 

\begin{figure}[h]
\includegraphics[width=\columnwidth]{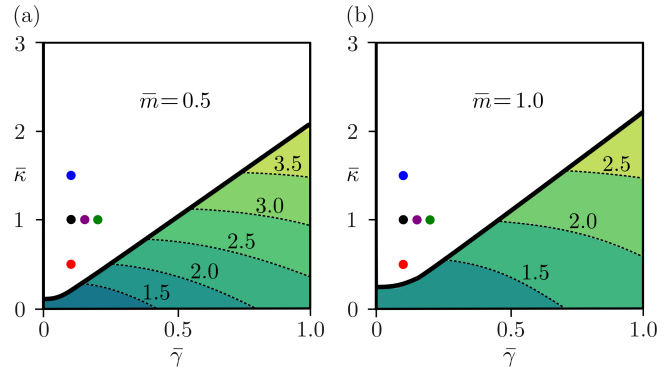}
\caption{(Color online) Diagram illustrating the existence of isolated frequencies of the matrix $\di W_+$ for different coupling strengths~$\di\gamma$ and~$\di\kappa$ and the two mass ratios $\di m=0.5$ (a) and $\di m=1$~(b). At least one isolated frequency is found for the parameters below the black boundary line: The value of the largest one is indicated by the contour lines. The dots in the white region indicate the parameter values used in our simulations: They all lie in the region where no isolated frequencies exist.}
\label{fig5}
\end{figure}

Here, we are interested in the parameter regime in which the COM motion of the defect reaches a stationary state before the time scale $\trev$, and in particular, when this stationary state is a thermal state at the temperature $T$ in which the chain was initially prepared. In order to identify the coupling strengths~$\di \gamma$ and~$\di \kappa$ for which this is verified, we perform a numerical search of the isolated frequencies of the matrix $\di W_+$  The results are presented in Figs.~\ref{fig5} for mass ratios $\di m=0.5$ and $\di m=1$. In the region above the broad curves no isolated frequency of $\di W_+$ was found. In this domain thermalization of the defect COM occurs in the thermodynamic limit according to our numerical simulations. In the segmented region below the boundary curve, at least one isolated frequency of $\di W_+$ exists. Here, the labels on the contour lines indicate the value of the largest isolated frequency of the normal modes. The dots in the upper region indicate the parameter values used in our simulations: They all lie in the region where no isolated frequencies exist.

\subsection{Finite chains}

Since our analysis is essentially numerical, we consider finite chains and we aim at observing a (transient) stationary state of the COM motion before finite size effects become relevant. The latter are characterized by the time scale ${\trev= L/ c_{\rm s}}$, where $ L=N  a$ is the chain length and $ c_{\rm s}=\omcut\, a/2$ the sound velocity \cite{footnote:trev}. The time scale $\trev$ grows linearly with the particle number~$N$, showing that by choosing a sufficiently large number of particles thermalization of the defect particle could be observed before finite-size effects become significant (which we denote by ``revivals''). 

\begin{figure}[h] 
\includegraphics[width=0.78\columnwidth]{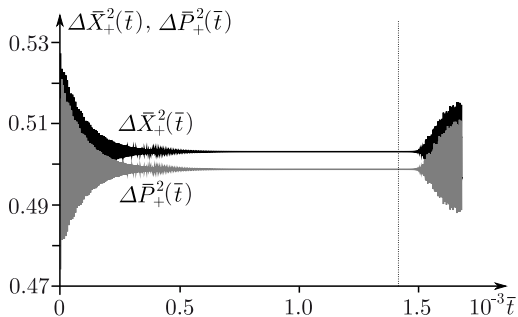}
\caption{Time evolution of the variances~$\DXsquare(\di t)$ (black) and~$\DPsquare(\di t)$ (grey) for the temperature $\di T=10^{-5}$ of the chain. After a transient time, the covariance matrices reach the stationary values $\DXsquare(\di T)=0.5031$ and $\DPsquare(\di T)=0.4988$, which indicate that the harmonic chain causes squeezing of the momentum variable at very low temperatures. The dotted vertical line represents the revival time $\tth\approx 1416$. The other parameters were chosen to be $\di m=0.5$, $\di \kappa=1$, $\di \gamma=0.1$, $\di\phi_1=\di \phi_2=0$, and $\di r_1=\di r_2=\frac14\ln(1-\di \gamma)$.
}
\label{fig5b} 
\end{figure}

We illustrate the thermalization of the COM motion of the defects by showing the time evolution of the variances~${\DXsquare(\di t)=\ave{\di X^2_+(\di t)}-\ave{\di X_+(\di t)}^2}$ and ${\DPsquare(\di t)=\ave{\di P^2_+(\di t)}-\ave{\di P_+(\di t)}^2}$ in Fig.~\ref{fig5b}. After a transient time, the variances approach their stationary values~${\DXsquare(\di T)}$ and~${\DPsquare(\di T)}$ that depend on the initial temperature of the chain, but not on the initial squeezing of the defects. The appearance of revivals after~$\trev$ is clearly visible.

A good estimate for~${\DXsquare(\di T)}$ and~${\DPsquare(\di T)}$ follows from the assumption that the total system (defects and harmonic chain) is in the thermal state ${\di \rho_{\rm th}=\exp(-\di\beta \di H)/\di Z_{\rm th}}$. Here, $\di H$ denotes the total Hamiltonian~\eqref{eq:Hdim}, $\di \beta$ the inverse temperature of the initial chain~\eqref{eq:InitialStateChain}, and $\di Z_{\rm th}={\rm Tr}\{\exp(-\di\beta \di H)\}$ the corresponding partition function. With the help of Eqs.~\eqref{eq:CovarianceBathdim} and~\eqref{eq:PotMatrixVPlus}, we find the following values for the stationary state variances of the defects:
\begin{equation*}
\begin{split}
\DXsquare(\di T)&=\frac{1}{2}\left[\di W_+^{-\frac{1}{2}}\,
\coth\bigg(\frac{\di \beta}{2}\di W_+^{\frac12}\bigg)\right]_{11}\,, \\
\DPsquare(\di T)&=\frac{1}{2}\left[\di W_+^{\frac{1}{2}}\,
\coth\bigg(\frac{\di \beta}{2}\di W_+^{\frac12}\bigg)\right]_{11}\,.
\end{split}
\end{equation*}
The indices on the right-hand side of the equations indicate the $(1,1)$-elements of the matrices inside the brackets. The fact that this estimate works so well, despite the unitary time evolution of the total system, is reminiscent of the concept of ``canonical typicality''~\cite{Winter,goldstein2006} that recently gained a lot of attention.

The parameter values of our microscopic model have to meet several constraints. First of all, the coupling strength $\di \kappa$ must be sufficiently large in order to guarantee that the frequency of the two oscillators lies well below the cutoff frequency $\omcut$ and more specifically in the linear region of the spectral density. Moreover, the value $\di \gamma$ must be sufficiently small such that the dispersion spectrum of the harmonic chain is not significantly perturbed by the coupling with the defects. There is also a further bound to the coupling strength~$\di \gamma$ that stems from the necessity to reduce computational resources. In fact, $\di\gamma$ determines the rate at which the center-of-mass motion reaches a stationary state. Very small values of $\di\gamma$ would require that one chooses an increasing particle number~$N$ in order to observe a stationary state well before~$\trev$, which results in a formidable computational problem. 

In order to account for all these requirements, we have chosen the parameter values $\di m=0.5$, $\di \gamma=0.1$, and $\di \kappa=1$ as standard parameters for our numerical simulations throughout this paper. As in the previous figures, we illustrate the changes in the numerical results that arise from different coupling constants by using the two parameter sets: {\it (i)~The $\di \gamma$-variation parameters.} The results are presented for three different coupling strengths ${\di\gamma=0.1,0.15,0.2}$ and for the fixed parameter value $\di\kappa=1$. {\it(ii)  $\di \kappa$-variation parameters.} The results are presented for constant~$\di\gamma=0.1$ and for variable $\di\kappa=0.5,1,1.5$. For the case $\di\gamma=0.1$, $\di\kappa=1.5$ we used $N=2000$ ions in the chain, while in all other cases it was sufficient to work with $N=1000$ ions in order to observe that the COM motion reached a stationary state well before the revival time~$\trev$. 

\begin{figure}[h] 
\includegraphics[width=\columnwidth]{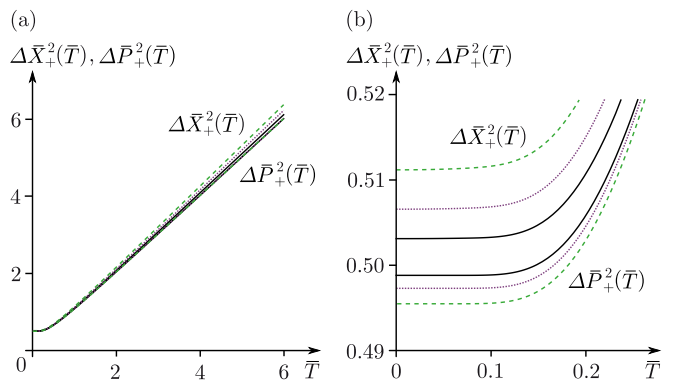}
\caption{(Color online) Variances $\Delta \di X_+^2(\di T)$ (three upper curves) and $\Delta \di P_+^2(\di T)$ (three lower curves) of the COM motion after thermalization for large temperatures $\di T\in[0,6]$ (a) and for low temperatures $\di T\in[0,0.2]$ (b). The parameters are $\di m=0.5$, $\di\kappa=1$, and  $\di\gamma=0.1$ (solid), $\di\gamma=0.15$ (dotted), and $\di\gamma=0.2$ (dashed line).}
\label{fig7} 
\end{figure}

It is instructive to analyze the variances of the COM position and momentum, after the steady state has been reached. Figure~\ref{fig7} shows the variances~$\DXsquare(\di T)$ and $\DPsquare(\di T)$ for a large temperature range $\di T\in [0,6]$ (a) and for small temperatures $\di T\in [0,0.2]$ (b) given the $\di\gamma$-{\it variation parameters} with $\di m=0.5$. For large temperatures (a) the variances grow linearly with a slightly different slope for the individual coupling strengths~$\di \gamma$, whereby~${\DXsquare(\di T)>\DPsquare(\di T)}$. In the low temperature regime (b), we observe squeezing of the COM momentum ${\DPsquare(\di T)<1/2}$ that increases for larger coupling strengths~$\di \gamma$. We note that this squeezing is induced by the coupling with the bath and has been identified in the studies reported in \cite{paz2008,paz2009}. It is reminiscent of the squeezing found for large coupling strengths in the Drude model~\cite{grabert1984,weiss1999}.

\begin{figure}[h] 
\includegraphics[width=\columnwidth]{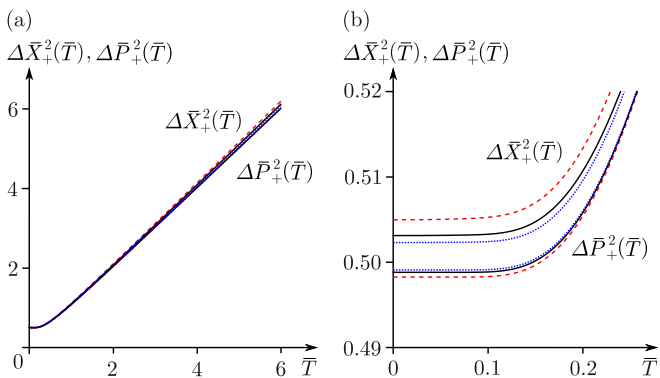}
\caption{(Color online) Variances $ \Delta\bar X_+^2(\bar T)$ (three upper curves) and $\Delta\bar P_+^2(\bar T)$ (three lower curves) for large and low temperatures, $\di T\in[0,6]$ (a) and $\di T\in[0,0.2]$ (b), respectively. The parameters are $\di m=0.5$, $\di\gamma=0.1$, and $\di\kappa=0.5$ (dashed), $\di\kappa=1$ (solid), and $\di\kappa=1.5$ (dotted). }
\label{fig8}
\end{figure}

Figure~\ref{fig8} displays the corresponding behavior of the COM variances for the $\di \kappa$-{\it variation parameters}. As before, we find a linear behavior of $\Delta \bar X_+^2(\bar T)$ and $ \Delta\bar P_+^2(\bar T)$ for large temperatures (a). For low temperatures (b), the squeezing of the variance $\Delta\di P_+^2(\di T)<1/2$ becomes larger for smaller~$\di \kappa$ and vice versa.

Finally, we emphasize that a mass ratio~$\di m=1$ leads only to marginal changes in the temperature behavior of the variances $\Delta\di X_+^2(\di T)$ and $\Delta\di P_+^2(\di T)$. 

These properties directly affect the behavior of bath-mediated entanglement between the defect oscillators, as we show in Sec.~\ref{sec:LogarithmicNeg}.

\subsection{Memory effects}

We now analyze memory effects in our model using our parameter choice. For this purpose we discuss the memory-friction kernel $\di\Gamma_+(\di t)$ of the GQLE~\eqref{eq:GQLEModel1} that is connected to the spectral density $\di J_+(\di\omega)$ for $\di t\geq 0$ according to relation
\begin{equation*}
\di\Gamma_+(\di t)= \frac{2}{\pi}\int_{0}^\infty \frac{\di J_+(\di \omega)}{\di \omega}\,\cos(\di\omega\di t)\,d\di\omega\,,
\end{equation*}
which inverts Eq. \eqref{eq:SpectralDensity}. For strict Ohmic dissipation, the memory-friction kernel would read $\di\Gamma_+(\di t)=2 \di \Gamma\, \delta(\di t)$ with $\di \Gamma$ as friction constant, and the GQLE would reduce to the ordinary Langevin equation without memory effects, provided that the ``slip term'' $-2 \Gamma\, \delta(\di t)\,\di X_+(0)$ and the oscillator frequency shift $-2 \Gamma\, \delta(0)\,\di X_+(\di t)$ can be disregarded \cite{weiss1999}. However, in our model we do not meet the requirements of a strict Ohmic environment since the cutoff frequency $\di \omega_{\rm cut}$ is only a few times larger than the oscillator frequency of the two coupled oscillators. For this reason, memory effects are present. 
\begin{figure}[h]
\includegraphics[width=\columnwidth]{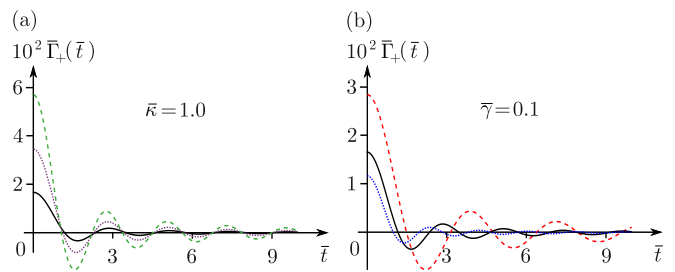}
\caption{(Color online) Memory-friction kernel $\di \Gamma_+(\di t)$. The plots in (a) and (b) are evaluated for the parameters of  the curves for the spectral density in Fig. \ref{fig3} (a) and (b), respectively.}
\label{fig4}
\end{figure}

The figures \ref{fig4}(a) and (b) display the memory-friction kernel as a function of time: An oscillatory decay is observed over a time scale~$\di t$ that is of the order of one, corresponding to $t\sim 1/\Omega_\gamma$ \cite{footnote:decay}. Hence, non-Markovian effects are present, but irrelevant for the dynamics of entanglement generation between the defects, as is shown in the following.

\section{Chain-mediated entanglement between the defects\label{sec:LogarithmicNeg}}

In this section we report the predictions of our model on correlations between the two defect oscillators. Entanglement is quantified by means of the logarithmic negativity~\cite{vidal2002,plenio2005}, that is evaluated using the covariance matrix at the time scale at which the COM motion of the two defects has reached a (quasi) steady state (before the revival time). We present the results for the logarithmic negativity found for different choices of parameters, such as the initial squeezing of the defect oscillator, the temperature of the chain, and the coupling strength between chain and defects.


\subsection{Logarithmic negativity of the oscillators\label{Sec:LogNegVariancesModel1}}

Since the state of the two defects is a Gaussian state at all times, the most convenient entanglement measure for our purpose is the logarithmic negativity~\cite{vidal2002,plenio2005,footnote:logneg}.

In what follows, we present the final results and refer to Appendix~\ref{App:LogNeg} for further details on the calculations. The logarithmic negativity reads
\begin{equation}
E_{N}(\tth)=\op{max}\left\{0,\mathcal{E}_N(\tth)\right\}
\label{eq:LogNeg}
\end{equation}
and contains the function ${\mathcal{E}_N(\tth)=-\ln(2\,\tilde \nu_-(\tth))}$, which depends on the smallest symplectic eigenvalue~$\tilde\nu_{-}(\tth)$ of the partial transpose of the covariance matrix $\di\Sigma(\tth)$, defined in Eq. \eqref{eq:sigma:t}. The covariance matrix $\di\Sigma(\tth)$ describes the state of the system for sufficiently long times, $\di t\sim \tth$, after which the COM motion has reached a stationary state independent of its initial state. The smallest symplectic eigenvalue~$\tilde\nu_{-}(\tth)$ follows from the identity~\cite{adesso2005,pirandola2009}
\begin{equation}
\tilde\nu_{-}(\tth)=\frac{1}{\sqrt{2}}\left(\tilde \Delta(\tth)-\sqrt{\tilde\Delta^2(\tth)-4\det \di\Sigma(\tth)}\,\right)^{\!\frac12}\,,
\label{eq:tildenu}
\end{equation}
with the time-independent determinant 
\begin{align}
\det\di\Sigma(\tth)=&\,\frac18\,\DXsquare\,\DPsquare\Big(1+\cosh(2\di r_1)\cosh(2\di r_2)\Big.
\nonumber\\
\Big.&-\cos(\Delta\di\phi)\sinh(2\di r_1)\sinh(2\di r_2)\Big)\,,
\label{eq:DeterminantMod1}
\end{align}
and the oscillating auxiliary function 
\begin{equation}
\tilde \Delta(\tth)=\tilde \Delta_0+\tilde \Delta_2 \cos(2 \tth + \di\varphi)\,.
\label{eq:tildeDeltat}
\end{equation}
In the last two expressions, we have introduced the relative squeezing angle ${\Delta\di\phi=\di\phi_2-\di\phi_1}$, as well as the coefficients 
\begin{equation}
\tilde \Delta_0=\frac14\Big(\DXsquare+\DPsquare\Big)\Big(\cosh(2\di r_1)+\cosh(2\di r_2)\Big)
\label{eq:DeltaZero}
\end{equation}
and 
\begin{align}
\tilde \Delta_2= & \, \frac14\,\Big|\DXsquare-\DPsquare\Big|\Big(\sinh^2(2\di r_1)+\sinh^2(2\di r_2)\Big.
\nonumber\\
\Big.&+2\,\cos(\Delta\di\phi)\,\sinh(2\di r_1)\,\sinh(2\di r_2)\Big)^{\!\frac12}\,.
\label{eq:Deltatwo}
\end{align}
The constant phase $\di\varphi$ can be determined, but is of no further interest to us. Due to the periodicity of the auxiliary function~\eqref{eq:tildeDeltat} the quantity ${\mathcal{E}_N(\tth)}$ oscillates for~${\tth<\trev}$ between a minimal and maximal value $\mathcal{E}^{\rm min}_N$ and $\mathcal{E}^{\rm max}_N$.
The formulas~\eqref{eq:LogNeg}-\eqref{eq:Deltatwo} provide a generalization of previously obtained expression for the logarithmic negativity~\cite{paz2008,paz2009}.

\begin{figure}[h] 
\includegraphics[width=\columnwidth]{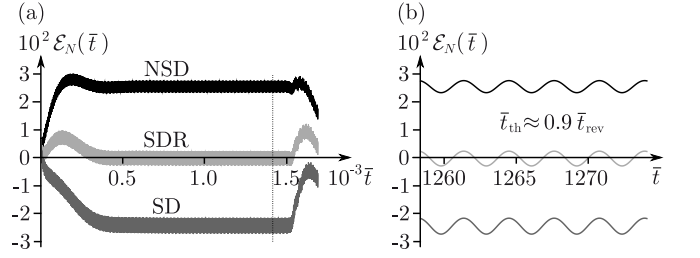}
\caption{Time evolution of the logarithmic negativity~$\mathcal{E}_N(\di t)$ for three different temperatures ${\di T=10^{-5}},0.27,0.33$ (from top to bottom). The curves exemplify the different behaviors of the entanglement (NSD, SDR, SD).  The dotted vertical line in (a) represents the revival time $\trev$.  (b)~Detail of the behavior about the time $\di t\approx 0.9\,\trev$. The other parameters are $\di m=0.5$, $\di \kappa=1$, $\di \gamma=0.1$, $\di\phi_1=\di \phi_2=0$ and $\di r_1=\di r_2=\frac14\ln(1-\di \gamma)$. Here, $\trev\approx 1416$.}
\label{fig6} 
\end{figure}

Following the nomenclature of~\cite{paz2008,paz2009}, we distinguish three qualitatively different situations for the entanglement of the two oscillators. (i)~When $\mathcal{E}^{\rm max}_N<0$, the logarithmic negativity is zero and we find no entanglement between the oscillators. We call this scenario the {\it sudden death}~(SD) phase because any transient entanglement disappears abruptly before the thermalized state is reached. (ii)~When ${\mathcal{E}^{\rm min}_N<0<\mathcal{E}^{\rm max}_N}$, we obtain an alternating sequence of periods with zero and nonzero logarithmic negativity, the so-called {\it sudden death and revival}~(SDR) phase. (iii) Finally, when $\mathcal{E}^{\rm min}_N>0$ the two oscillators are entangled after thermalization which we call the {\it no sudden death}~(NSD) phase. In Fig.~\ref{fig6} we exemplify these different phases by showing the time evolution of~$\mathcal{E}_N(\tth)$ for three initial temperatures. Figure~\ref{fig6}(a) displays the long-time behavior of~$\mathcal{E}_N(\tth)$, its evolution toward the steady state. Here, the occurrence of revivals after~$\trev$ are visible. Figure~\ref{fig6}(b) zooms in the behavior at~$\tth\approx 0.9 \,\trev$, showing that the logarithmic negativity exhibits oscillations at the frequency $\Omega_\gamma$. These oscillations have been also observed in Refs. \cite{paz2008,paz2009} and their physical origin simply lies in the decoupling of the relative coordinate from the rest of the dynamics. In fact, the squeezed variance of the relative motion rotates with frequency $2\Omega_\gamma$, and correspondingly the smallest symplectic eigenvalue oscillates at the same frequency. We also note that, by choosing the squeezing parameters according to $\di r_1=\di r_2=0$, we find by virtue of Eqs.~\eqref{eq:tildeDeltat} and~\eqref{eq:Deltatwo} that the logarithmic negativity~\eqref{eq:LogNeg} of the steady state becomes time independent and displays no further oscillations. The underlying reason is that for ${\di r_1=\di r_2=0}$ the initial state of the relative motion corresponds to the ground state of the Hamiltonian~\eqref{eq:Hminus}.

In this context, we would like to point out that the NSD phase can be characterized by the fulfillment of the inequality (see also Eq.~\eqref{eq:EntanglementCondition} in Appendix~\ref{App:LogNeg})
\begin{equation}
 \tilde \Delta_0-\tilde \Delta_2-4\,\det\di\Sigma(\tth)-\frac14>0\,,
\label{eq:EntanglementCriterion}
\end{equation}
which follows from ${\mathcal{E}_N(\tth)=-\ln(2\,\tilde \nu_-(\tth))>0}$ or equivalently $\tilde \nu_-(\tth)<\frac12$ evaluated for the minimal value of $\tilde \Delta(\tth)$, Eq.~\eqref{eq:tildeDeltat}. Thus, if inequality~\eqref{eq:EntanglementCriterion} is satisfied, the two defect oscillators are entangled after the COM has reached a stationary state (before the revival time~$\trev$). This inequality in connection with the identities~\eqref{eq:DeterminantMod1}-\eqref{eq:Deltatwo} provides a general criterion for the existence of steady-state entanglement for arbitrary initial squeezed states of the defects.

\subsection{Entanglement generation for different initial parameters and coupling strengths\label{Sec:LogNegParameterChangesFirstModel}}

In this section we report the logarithmic negativity of the defect oscillators after the COM defect oscillator has reached a stationary state, for different values of the initial squeezing of the defects and of the initial temperature of the ion chain. The results are displayed using the type of contour plots first introduced in Ref.~\cite{paz2008,paz2009}, which highlight the different entanglement regions (NSD, SD and SDR) as a function of the modulus of the initial squeezing parameter and the temperature of the reservoir.

We first consider the case in which the initial states of the defect oscillators are characterized by the same squeezing parameters,~$\di r_1=\di r_2=\di r$ and $\Delta\di \phi=0$.  We use the inequality~${\DXsquare(\di T)>\DPsquare(\di T)}$, which we found numerically for the considered parameter regime, and reduce the determinant~\eqref{eq:DeterminantMod1} to the form
\begin{equation}
\det\di\Sigma(\tth)=\,\frac14\,\DXsquare\,\DPsquare\,,
\label{eq:detSigmaPhiZero}
\end{equation}
while the coefficients~\eqref{eq:DeltaZero} and~\eqref{eq:Deltatwo} read
\begin{align}
\tilde \Delta_0= &\frac12\,\Big(\DXsquare+\DPsquare\Big)\cosh(2\di r)
\label{eq:DeltaZeroPhiZero}\\
\tilde \Delta_2= &\frac12\,\Big(\DXsquare-\DPsquare\Big)\sinh(2\di r)\,.
\label{eq:DeltaTwoPhiZero}
\end{align}
These expressions lead to the following simple form of the entanglement condition~\eqref{eq:EntanglementCriterion} that characterizes the NSD phase:
\begin{equation*}
\frac12\Big(\DPsquare\, e^{2\di r} + \DXsquare\, e^{-2\di r}\Big)-\DXsquare\,\DPsquare-\frac14>0\,.
\end{equation*}
With the substitution $y=e^{2\di r}$, this relation reduces to a quadratic inequality in $y$ that yields two independent conditions for the steady-state entanglement of the two oscillators. The first of these conditions reads
\begin{equation}
\DXsquare(\di T)<\frac12\,e^{2\di r}
 \label{eq:EntanglementCondMod1I}
\end{equation}
and tells us that entanglement between the oscillators will occur at any temperature~$\di T$ as long as the initial squeezing parameter $\di r$ is sufficiently large. The underlying mechanism for this entanglement generation is based on the existence of a decoherence free subspace, following from the decoupling of the relative motion.

The second entanglement condition takes the form
\begin{equation}
\DPsquare(\di T)<\frac12\,e^{-2\di r}
 \label{eq:EntanglementCondMod1II}
\end{equation}
and is only satisfied for sufficiently small squeezing parameters~$\di r$ and temperatures $\di T$. We call this second mechanism {\it bath-induced entanglement} because it arises from the squeezing of~$\DPsquare(\di T)$ caused by the interaction of the oscillators with the reservoir. It is clear that the two mechanisms are competing.

Figure~\ref{fig9} displays the different phases of entanglement for varying~$\di r$ and~$\di T$ including the contour lines of the logarithmic negativity within the NSD region. In Fig. \ref{fig9}(a) one can observe the behavior at large squeezing and high temperatures. Here,  entanglement in the NSD region is due to the decoupling of the relative motion and is determined by the condition~\eqref{eq:EntanglementCondMod1I}. The SDR region is not visible, but lies between the NSD and SD phases. Figure \ref{fig9}(b) displays the behavior at small squeezing and low temperatures. One can here observe the NSD island, which occurs in the vicinity of $\di r=0$, $\di T=0$ and is separated by the SDR phase from the main NSD region. This island stems from the bath-induced entanglement according to Eq.~\eqref{eq:EntanglementCondMod1II}. 

\begin{figure}[h] 
\includegraphics[width=\columnwidth]{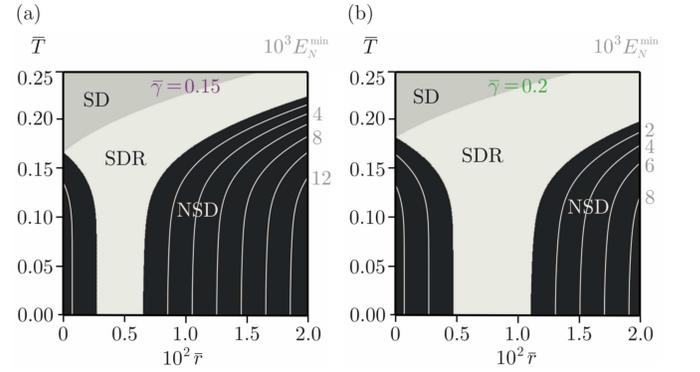}
\caption{Contour plots of the logarithmic negativity~$E_N(\di r,\di T)$ for $\di m=0.5$, $\di \kappa=1$, and (a) $\di \gamma=0.15$, (b) $\di \gamma=0.2$.}
\label{fig10} 
\end{figure}

\begin{figure}[h] 
\includegraphics[width=\columnwidth]{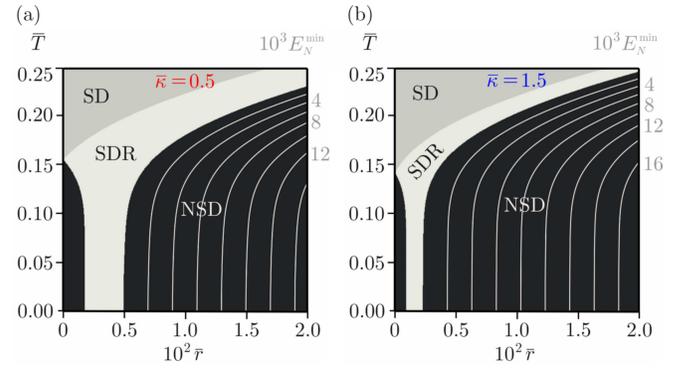}
\caption{Contour plots of~$E_N(\di r,\di T)$ for $\di m=0.5$, $\di\gamma=0.1$, and (a) $\di \kappa=0.5$, (b) $\di \kappa=1.5$.}
\label{fig11} 
\end{figure}

Since the squeezing of the COM motion at low temperatures is rather small, the NSD region due to bath-induced entanglement covers only a small region of Fig.~\ref{fig9}(b). The size of the region can be increased by increasing the squeezing of the variance $\DPsquare(\di T)$. According to Fig.~\ref{fig7}(b), this can be achieved by increasing the parameter $\di \gamma$. Figure~\ref{fig10} displays the corresponding contour plots in the regime of small squeezing parameters and low temperatures for two values of the coupling strength~$\di \gamma$: An increase of the NSD region of bath-induced entanglement is observed for larger coupling strengths~$\di \gamma$. We recall, however, that this behavior can saturate, when $\di \gamma$ takes values at which the transient steady state is not reached before $\trev$. The squeezing of the COM variance can also be increased by decreasing the coupling strength~$\di \kappa$, as illustrated in Fig.~\ref{fig8}(b). Figure~\ref{fig11} depicts the change in the entanglement behavior for varying $\di \kappa$. Here one can see that the size of the region where bath-induced entanglement is found is larger for smaller values of $\kappa$. 

When the two oscillators are instead prepared in squeezed states with a relative squeezing angle $\Delta\di\phi\neq 0$,, the entanglement will be diminished. In fact, such initial states lead to a smaller squeezing of the relative motion. A representative situation is found for $\Delta\di\phi=\pi$, namely, when the squeezed quadratures of the defect oscillators are orthogonal. In this case the relative motion is not squeezed and one obtains for the determinant~\eqref{eq:DeterminantMod1} and the coefficients~\eqref{eq:DeltaZero} and~\eqref{eq:Deltatwo}
with~$\di r_1=\di r_2=\di r$ the expressions
\begin{equation*}
\det\di\Sigma(\tth)=\,\frac14\,\DXsquare\,\DPsquare\,\cosh^2(2\di r)\,,
\end{equation*}
as well as
\begin{equation*}
\tilde \Delta_0= \frac12\Big(\DXsquare+\DPsquare\Big)\cosh(2\di r)\quad\text{and}\quad \tilde \Delta_2= 0\,.
\end{equation*}
In this case, the entanglement condition~\eqref{eq:EntanglementCriterion} reduces to 
\begin{equation*}
\frac12\Big(\DXsquare+\DPsquare\Big)\cosh(2\di r)-\DXsquare\,\DPsquare\,\cosh^2(2\di r)-\frac14>0\,.
\end{equation*}
This inequality is fully equivalent to the new criterion
\begin{equation*}
\DPsquare(\di T)<\frac{1}{2\,\cosh(2\di r)}\,,
\end{equation*}
which is only satisfied for a squeezed COM momentum, in analogy to Eq.~\eqref{eq:EntanglementCondMod1II}. It shows that entanglement between the defects can only be generated by the active coupling with the bath. The existence of a decoherence free subspace does not support entanglement generation in this case.

Thus, the relative squeezing angle ${\Delta\di\phi}$ can be used as a control parameter to distinguish between the two mechanisms that lead to steady-state entanglement. This observation makes our model a favorable microscopic setting to study the generation of bath-induced entanglement.


\section{Conclusions}
\label{Sec:Conclusions}

We have numerically investigated the dynamics of two defects coupled to one edge of a harmonic crystal and identified the parameter regime for which the defect variables reach a quasi steady state.
This (quasi) equilibrium sets in for time scales which are smaller than the revival time scale characterized by finite size effects. From its features and its scaling behavior for different system sizes, we can conclude that it corresponds to the equilibrium reached in the thermodynamic limit, when the number of ions of the chain is infinitely large. The analysis of the correlations between the defects shows that they can become entangled in the steady state. Such entanglement emerges as a consequence of the symmetries of the Hamiltonian, and it follows the dynamics outlined in Refs.~\cite{paz2008,paz2009} where it was determined by means of an effective master equation mimicking the effect of the bath. Our work shows that a physical model, such as the considered extension of Rubin's model, establishes a microscopic realization of this dynamics. It allows us to determine the relevant time scales which emerge from the spectral properties of the chain, the defects, and their mutual coupling. Moreover, it gives us the possibility to analyze the dynamics in regimes where a master equation approach is not convenient (e.~g. when finite size effects become relevant).

This work provides a microscopic understanding of the dynamics of bath induced entanglement, building upon the general criterion given by Eqs.~\eqref{eq:LogNeg}-\eqref{eq:EntanglementCriterion}. Based on a realistic model, it goes beyond the reach of idealized settings studied so far that employ ideal bosonic heat baths with artificially chosen spectral densities. An interesting next step will be the extension of our model to non-Gaussian initial states and nonquadratic Hamiltonians for the defects. As long as the symmetry is preserved, we anticipate that the underlying mechanisms will support the formation of steady-state entanglement. Whether such an extension will lead to an enhancement in the entanglement generated between the defects is however an open question.

In a follow up to this article we will discuss the generation of entanglement between two defects that couple to distant sites of the chain, thereby extending and complementing the findings reported in Ref.~\cite{wolf2011}, which were not addressed in the present \mbox{article}.

\begin{acknowledgments}
We thank C.~Cormick, G. De Chiara, T. Fogarty, M.~B.~Plenio, W.~P.~Schleich, and B. Taketani for fruitful discussions. We acknowledge financial support of the European Commission (Integrating Project AQUTE; STREP PICC), of the German Research Foundation (LU1382/1-1 and Heisenberg program), of the BMBF QuORep, of the  Spanish Ministerio de Ciencia y Innovaci{\'o}n (Acci{\'o}n Complementaria, EUROQUAM "CMMC: Cavity Mediated Molecular Cooling"), and of the cluster of excellence ``Nanosystems Initiative Munich (NIM)''.
\end{acknowledgments}


\appendix

\section{Transformation of the squeezing parameters\label{App:NewSqueezingParameters}}

Based on the dimensionless description of Subsec.~\ref{Sec:DimensionlessDes} and the matrices $S(z)$ and $O(\phi)$, Eq.~\eqref{eq:DefOandS}, we find for the initial covariance matrices of the defects~$\sigma_\mu(0)$, Eqs.~\eqref{eq:sigma11}-\eqref{eq:sigma12}, the dimensionless form
\begin{equation}
\di \sigma_\mu(0)=\frac12 \,S(\di\Omega^{\frac12})\,O^\tp(\phi_\mu)\,
S(e^{2 r_\mu})\,O(\phi_\mu)\,S(\di\Omega^{\frac12})\,.
\label{eq:CovarianceSystemdim}
\end{equation}
Due to the outer symplectic matrices $S(\di \Omega)$, we would arrive at much more complicated expressions for the logarithmic negativity in Sec.~\ref{Sec:LogNegVariancesModel1} when starting from Eq.~\eqref{eq:CovarianceSystemdim}. These expressions would conceal the class of squeezing parameters that lead to the same steady-state entanglement between the defects. 

For this reason, it is advantageous to introduce new squeezing parameters $\di s_\mu=\di r_\mu e^{{\rm i}\di\phi_\mu}$ that overcome these difficulties by transforming the covariance matrices~\eqref{eq:CovarianceSystemdim} to the simpler form~\eqref{eq:CovarianceSystemdimStandard}.
The corresponding transformation equations $r=r(\di r,\di \phi)$ and ${\phi=\phi(\di r,\di \phi)}$ follow directly from the diagonalization of~\eqref{eq:CovarianceSystemdim} and a subsequent comparison of the resulting eigenvalues and eigenvectors with Eq.~\eqref{eq:CovarianceSystemdimStandard}.

In this way, we find that the one-to-one mapping between the new squeezing parameters ${\di r_\mu\geq 0}$, ${\di \phi_\mu\in (-\pi,\pi]}$ and the original ones (${r_\mu\geq 0}$, ${\phi_\mu\in (-\pi,\pi]}$) depends on~$\di\Omega$ and splits into three different domains of definition. 
Since we have $\di \Omega<1$, the mapping $r=r(\di r,\di \phi)$ and ${\phi=\phi(\di r,\di \phi)}$ reads for the special case $\di \phi=0$ ($\di r\geq 0$)
\begin{subequations}
\begin{align}
r(\di r,0)&=\textstyle{\di r-\frac{1}{2}\ln\di\Omega}\,,
\label{eq:squeezingTrafoDef1}\\
\phi(\di r,0)&=0\,.
\nonumber
\end{align}
For $\di \phi=\pi$ ($\di r>0$) we find accordingly
\begin{align}
r(\di r,\pi)&=
\textstyle{\left(\di r+\frac{1}{2}\ln\di\Omega\right)\cdot\mathrm{sign}
\left(\di r+\frac{1}{2}\ln\di\Omega\right)}\,,
\label{eq:squeezingTrafoDef2}\\
\phi(\di r,\pi)&=\pi\cdot
\textstyle{\Theta\left(\di r+\frac{1}{2}\ln\di\Omega\right)}\,,
\nonumber
\end{align}
where $\Theta(x)$ denotes the Heaviside step function. The mapping of the remaining open domain ${\di r>0}$, ${\di \phi\in (-\pi,\pi)\setminus\{0\}}$ onto ${r>0}$, ${\phi\in(-\pi,\pi)\setminus\{0\}}$ is finally given by the one-parameter family of coordinate transformations
\begin{align}
 r(\di r,\di \phi)&=\frac12\,\mathrm{arcosh}\left(\di R_+\right)\,,
 \label{eq:squeezingTrafoDef3}\\
\phi(\di r,\di \phi)&=2\, \mathrm{arctan}\!\left(\!
\frac{\sqrt{\di R^2_+-1}-\di R_- +\sinh(2\di r)\,\sin\di \phi}
{\sqrt{\di R^2_+-1}+\di R_- +\sinh(2\di r)\,\sin\di \phi}
\right)\,,
\nonumber
\end{align}%
\label{eq:squeezingTrafo}%
\end{subequations}%
with the auxiliary functions $\di R_\pm=\di R_\pm(\di r,\di \phi,\di\Omega)$ defined by
\begin{equation*}
\di R_\pm=\left(\frac{1}{\di\Omega}\pm\di\Omega\right)\frac{\cosh(2\di r)}{2}+
\left(\frac{1}{\di\Omega}\mp\di\Omega\right)\frac{\sinh(2\di r)}{2}\cos\di\phi\,.
\end{equation*}
Substitution of the transformation Eqs.~\eqref{eq:squeezingTrafo} into the original covariance matrix~\eqref{eq:CovarianceSystemdim} yields directly the convenient form~\eqref{eq:CovarianceSystemdimStandard}. 

The inverse transformation equations $\di r=\di r(r,\phi)$ and $\di \phi=\di \phi(r,\phi)$ follow in analogy to Eqs.~\eqref{eq:squeezingTrafo} by simply replacing the role of $\di \Omega<1$ in the derivation with its inverse $\di \Omega^{-1}>1$. Again, the domain of definition splits into three different parts. For the special case $\phi=0$ ($r\geq 0$) the inverse mapping is given by
\begin{subequations}
\begin{align}
\di r(r,0)&=
\textstyle{\left(r+\frac{1}{2}\ln\di\Omega\right)\cdot\mathrm{sign}\left(r+\frac{1}{2}\ln\di\Omega\right)}\,,
\label{eq:squeezingInverseTrafoDef1}\\
\di\phi(r,0)&=
\textstyle{\pi\left(1-\Theta\left(r+\frac{1}{2}\ln\di\Omega\right)\right)}\,.
\nonumber
\end{align}
For $\phi=\pi$ ($r>0$) it reads accordingly
\begin{align}
\di r(r,\pi)&=
\textstyle{r-\frac{1}{2}\ln\di\Omega}\,,
\label{eq:squeezingInverseTrafoDef2}\\
\di\phi(r,\pi)&=\pi\,.
\nonumber
\end{align}
As above, we find for the mapping of the open domain ${r>0}$, ${\phi\in(-\pi,\pi)\setminus\{0\}}$ onto  ${\di r>0}$, ${\di \phi\in (-\pi,\pi)\setminus\{0\}}$ a slightly more complicated expression
\begin{align}
\di r(r,\phi)&=\frac12\,\mathrm{arcosh}\left(\RR_+\right),
\label{eq:squeezingInverseTrafoDef3}\\
\di \phi(r,\phi)&=-2\, \mathrm{arctan}\!\left(\!
\frac{\sqrt{\RR^2_+-1}-\RR_- -\sinh(2r)\,\sin\phi}{\sqrt{\RR^2_+-1}+\RR_- -\sinh(2r)\,\sin\phi}\right),
\nonumber
\end{align}%
\label{eq:squeezingInverseTrafo}%
\end{subequations}%
with the auxiliary functions $\RR_\pm\!=\!\di\RR_\pm(r,\phi,\di\Omega^{-1})$ given by
\begin{equation*}
R_\pm=\left(\di\Omega\pm\frac{1}{\di\Omega}\right)\frac{\cosh(2r)}{2}+
\left(\di\Omega\mp\frac{1}{\di\Omega}\right)\frac{\sinh(2r)}{2}\cos\phi\,.
\end{equation*}
Using these inverse transformations, one can determine the values of the new squeezing parameters~$\di r_\mu$ and~$\di \phi_\mu$ for a given set of initial squeezing parameters $r_\mu$ and $\phi_\mu$.

\begin{figure}[h] 
\includegraphics[width=\columnwidth]{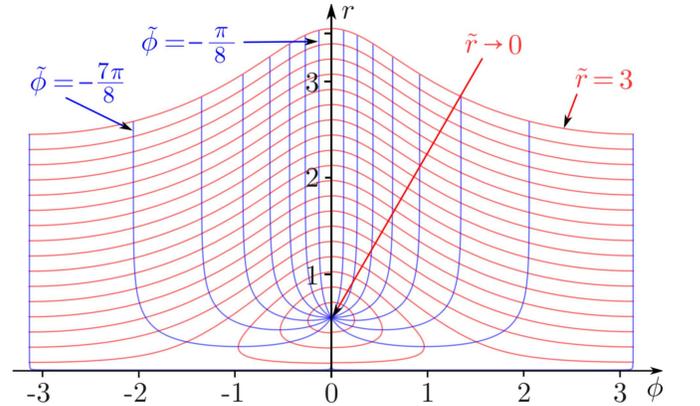}
\caption{(Color online) Illustration of the curvilinear coordinates defined by Eqs.~\eqref{eq:squeezingTrafo} for $\di\Omega=\frac{1}{3}$. The {\it red curves} indicate the coordinate lines $r=r(\di r,\di\phi)$ and $\phi=\phi(\di r,\di \phi)$ for constant $\di r>0$ and varying $\di\phi\in(-\pi,\pi)\setminus\{0\}$. The {\it blue curves} are obtained for constant~$\di \phi$ and varying~$\di r$.}
\label{fig15}
\end{figure}

The effect of the transformation Eqs.~\eqref{eq:squeezingTrafo} is illustrated in Fig.~\ref{fig15} by showing the coordinate lines $r=r(\di r,\di \phi)$ and $\phi=\phi(\di r,\di \phi)$ for $\di\Omega=\frac{1}{3}$ and constant values of $\di r$ and~$\di \phi$. The rather small value for~$\di\Omega$ was only chosen to highlight the effect of the transformation Eqs.~\eqref{eq:squeezingTrafo}, and does not correspond to any of the parameter values used throughout the paper.

We conclude this appendix by pointing out that the transformation Eqs.~\eqref{eq:squeezingTrafo} are only valid for ${\di\Omega<1}$. In the case of ${\di\Omega>1}$, one obtains the corresponding transformation equations by simply interchanging Eqs.~\eqref{eq:squeezingTrafo} and its inverse~\eqref{eq:squeezingInverseTrafo} and replacing 
$r\leftrightarrow \di r$ and $\phi\leftrightarrow \di \phi$.


\section{Spectral density for different trap frequencies $\di \omega_B$\label{App:SpectralDensitySpecialCase}}

The purpose of this appendix is to show that the shape of the spectral density depends crucially on the choice of the edge frequency $\di\omega_B$. In the main part of the paper, we restrict ourselves to the fixed value $\di\omega_B=\sqrt{\di\kappa/\di m}$. In this way, we compensate for the missing frequency shift of the ions at the end of the chain (they couple only to one neighboring ion). This choice yields a suitable tridiagonal form for the potential matrix~\eqref{eq:V} whose eigenvalues and eigenvectors can be analytically determined using the methods outlined in Ref.~\cite{Hu1996}. As a result, we find the spectrum in Eq.~\eqref{spectrum:0} for the specific trap frequency $\di\omega_B=\sqrt{\di\kappa/\di m}$.

\begin{figure}[h]
\includegraphics[width=\columnwidth]{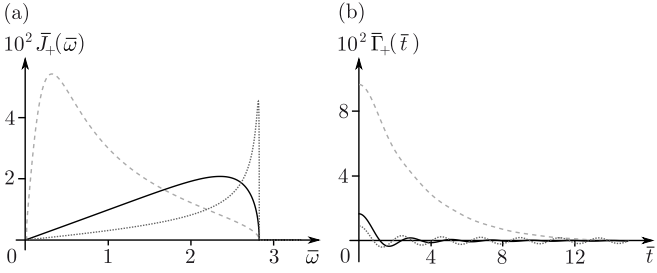}
\caption{Spectral density~(a) and the memory-friction kernel~(b) for the parameters $\di \gamma=0.1$, $\di \kappa=1$, $\di m=0.5$ and the three different trap frequencies~$\di \omega_B=0.1$ (dashed line), $\di \omega_B=\sqrt{2}$ (solid line), and $\di \omega_B=2$ (dotted line).}
\label{fig16}
\end{figure}

Since the two defect oscillators couple to the edge particle of the harmonic chain, the trap frequency~$\di\omega_B$ has an immediate influence on the behavior of the reservoir. In order to illustrate this fact, we show in Fig.~\ref{fig16}(a) the spectral density for the standard parameters $\di \gamma=0.1$, $\di \kappa=1$, $\di m=0.5$ and the three different trap frequencies~$\di \omega_B=0.1$ (dashed curve), $\di \omega_B=\sqrt{2}$ (solid curve), and $\di \omega_B=2$ (dotted curve). Whereas~$\di J_+(\di \omega)$ exhibits a pronounced non-Ohmic behavior for small trap frequencies~$\di\omega_B\ll \sqrt{2}$, it still displays a linear growth in the neighborhood of $\di \omega=1$ for $\di\omega_B>\sqrt{2}$.

Figure~\ref{fig16}(b) depicts the corresponding memory-friction kernel~$\di \Gamma_+(\di t)$. For~$\di \omega_B=0.1$ we find a nonoscillatory, slowly decaying function~$\di \Gamma_+(\di t)$ which indicates large memory effects in the GQLE~\eqref{eq:GQLEModel1}. For~$\di \omega_B=2$ we obtain an oscillatory behavior of the memory-friction kernel; however, the oscillations do not decay for large times. The reason for this behavior is the existence of an isolated frequency in the spectrum of $\di W_+$ which prevents the COM motion from thermalization, see Sec.~\ref{Sec:reservoirCharcterization}.

\section{Analytic expressions for the logarithmic negativity of the steady state\label{App:LogNeg}}

In this appendix we derive the analytic expressions used in Sec.~\ref{sec:LogarithmicNeg} for the evaluation of the logarithmic negativity. We first recall how to find the logarithmic negativity for a given covariance matrix in general~\cite{vidal2002,plenio2005,adesso2005}. We then rewrite this formalism in COM and relative coordinates and apply it to the specific covariance matrix of the defects after they reached the steady state. Finally, we sketch the derivation of the simple expressions~\eqref{eq:DeterminantMod1}, \eqref{eq:tildeDeltat}, \eqref{eq:DeltaZero} and \eqref{eq:Deltatwo} that provide the logarithmic negativity for arbitrary initial squeezing parameters $\di s_\mu$ and steady-state variances $\DXsquare(\di T)$ and $\DPsquare(\di T)$.

\subsection{Logarithmic negativity and covariance matrix in COM and relative coordinates}

We start by recalling the definition of the combined vector of the position and momentum operators for the two defect oscillators ${\bxi=(\di X_1,\di P_1;\di X_2,\di P_2)^\tp}$. The corresponding covariance matrix is given by the expression ${\di\Sigma_{\alpha\beta}=\frac12\ave{\xi_\alpha\, \xi_\beta + \xi_\beta\, \xi_\alpha}-\ave{\xi_\alpha}\ave{ \xi_\beta}}$ with $\alpha,\beta\in\{1,\ldots 4\}$. It can be rewritten in the block form,
\begin{equation}
\di\Sigma=\left(
\begin{array}{cc}
A\phantom{^\tp} & C \\
C^\tp & B
\end{array}
\right)\,,
\label{eq:CovarianceBlockForm}
\end{equation}
where $A, B\in \mathds{R}^2$, denote the covariance matrices of the first and second defects, respectively, and $C\in \mathds{R}^2$ characterizes the correlations between them.
Next, we define the partially transposed covariance matrix $\tilde \Sigma=\Lambda\, \di \Sigma\, \Lambda$ with the help of the diagonal matrix $\Lambda=\op{diag}(1,1,1,-1)$. The logarithmic negativity~\cite{vidal2002,plenio2005} can then be determined from the smallest symplectic eigenvalue $\tilde \nu_-$ of~$\tilde \Sigma$ and reads
\begin{equation}
E_{N}=\op{max}\left\{0,-\ln\left(2\,\tilde\nu_{-}\right)\right\}\,.
\label{eq:LogNegApp}
\end{equation}
We note that the symplectic eigenvalues of $\tilde \Sigma$ coincide with the common, positive eigenvalues of the matrix~${-i J_2 \tilde\Sigma}$, where $J_2$ is given by Eq.~\eqref{eq:DefJ}. 

It is possible to write down an explicit expression for the smallest symplectic eigenvalue~\cite{adesso2005,pirandola2009} and for this purpose, we introduce the function
\begin{equation*}
\Delta(\di \Sigma)=\det A+\det B + 2 \det C
\label{eq:Delta}
\end{equation*}
which is invariant under symplectic transformations.
By applying this function to the partially transposed covariance matrix~$\tilde \Sigma$, we obtain the auxiliary function 
\begin{equation}
\tilde \Delta=\Delta(\Lambda\, \di \Sigma\, \Lambda)=\det A+\det B - 2 \det C\,.
\label{eq:DeltaTilde}
\end{equation}
With this quantity at hand, the smallest symplectic eigenvalue of $\tilde \Sigma$ follows from the identity
\begin{equation}
\tilde\nu_{-}=\sqrt{\frac{1}{2}\left(\tilde \Delta-\sqrt{\tilde\Delta^2-4\det \di\Sigma}\right)}\,.
\label{eq:tildenuapp}
\end{equation}
Given the covariance matrix in block form~\eqref{eq:CovarianceBlockForm}, we thus determine the logarithmic negativity~\eqref{eq:LogNegApp} by evaluating the smallest symplectic eigenvalue~\eqref{eq:tildenuapp} with the help of the auxiliary function~\eqref{eq:DeltaTilde}. 

Entanglement between the two oscillators is only found when $\mathcal{E}_N=-\ln(2\,\tilde \nu_-)>0$ which is equivalent to $\tilde\nu_-<\frac12$. Using Eq.~\eqref{eq:tildenuapp}, one can show that this entanglement condition coincides with the Simon criterion~\cite{simon2000}
\begin{equation}
\tilde \Delta-4\det\di\Sigma-\frac14>0\,.
\label{eq:EntanglementCondition}
\end{equation}

Now, due to the decoupling of the relative coordinate of the two defect oscillators in our microscopic model, we seek for an expression of~$\tilde \nu_-$ that is based on the covariance matrix in COM and relative coordinates. 
For this reason, we define in analogy to above the combined vector for the COM and relative coordinates ${\bxi^\crc=(\di X_+,\di P_+;\di X_-,\di P_-)^\tp}$ and write ${\di\Sigma^\crc_{\alpha\beta}=\frac12\aven{\xi^\crc_\alpha\, \xi^\crc_\beta + \xi^\crc_\beta\, \xi^\crc_\alpha}-\aven{\xi^\crc_\alpha}\aven{ \xi^\crc_\beta}}$ for the corresponding covariance matrix with block form
\begin{equation}
\di\Sigma^\crc=\left(
\begin{array}{cc}
\phantom{(}  A^\crc  \phantom{)^\tp} & C^\crc \\
(C^\crc)^\tp & B^\crc
\end{array}
\right)\,.
\label{eq:CovarianceBlockFormCOM}
\end{equation}
With the transformation matrix
\begin{equation*}
 R=\frac{1}{\sqrt{2}}
\left(
\begin{array}{cc}
\mathds{1}_2 & \phantom{-}\mathds{1}_2 \\
\mathds{1}_2 & -\mathds{1}_2
\end{array}
\right)=R^\tp=R^{-1}\,,
\end{equation*}
the connection between the COM and relative coordinates and their corresponding covariance matrices reads
\begin{equation}
\bxi^\crc=R\, \bxi \quad\text{and}\quad
\di\Sigma^\crc=R^\tp\,\di\Sigma\,R\,.
\label{eq:CovarianceTransform}
\end{equation}
In order to rewrite the quantities that appear in the smallest symplectic eigenvalue~\eqref{eq:tildenuapp} in terms of the block matrices 
$A^\crc$, $B^\crc$ and $C^\crc$, we take advantage of the fact that the transformation matrix $R$ is symplectic. An immediate consequence of this observation is the validity of the identities
\begin{equation}
\det \di \Sigma=\det \di \Sigma^\crc
\label{eq:DeterminateCOM}
\end{equation}
and
\begin{equation}
\Delta(\di \Sigma)=\Delta(R\,\di\Sigma^\crc\,R^\tp)=\Delta(\di \Sigma^\crc)\,.
\label{eq:DeltaSigmaApp}
\end{equation}
With the help of Eq.~\eqref{eq:DeltaSigmaApp}, we easily find the relation
\begin{equation}
\tilde\Delta\!=\!\Delta(\di \Sigma^\crc)-\det\!\left[A^\crc\!-\!B^\crc+(C^\crc)^\tp\!-\!C^\crc\right].
\label{eq:tildeDeltaCOM}
\end{equation}
In conclusion, the smallest symplectic eigenvalue~\eqref{eq:tildenuapp}, as well as the entanglement condition~\eqref{eq:EntanglementCondition} can be directly determined from the covariance matrix in COM and relative coordinates~\eqref{eq:CovarianceBlockFormCOM} by means of the identities~\eqref{eq:DeterminateCOM} and~\eqref{eq:tildeDeltaCOM}.

\subsection{The covariance matrix after thermalization of the COM motion}

The manifestation of correlations between the defects is a direct consequence of the decoupling of the relative coordinates and the thermalization of the COM motion. This statement can be well illustrated my means of the covariance matrix of the defect oscillators. Initially, the covariance matrix of the two defects reads
\begin{equation*}
 \di\Sigma(0)=\left(
\begin{array}{cc}
\di\sigma_1(0) & 0\\
0 & \di\sigma_2(0)
\end{array}
\right)\,,
\end{equation*}
where the $\di\sigma_\mu(0)$ are given by Eq.~\eqref{eq:CovarianceSystemdimStandard}. The transformation to COM and relative coordinates via Eq.~\eqref{eq:CovarianceTransform} yields the covariance matrix
\begin{equation*}
  \di\Sigma^\crc(0)=\frac12\left(
\begin{array}{cc}
\di\sigma_1(0)+\di\sigma_2(0)\; & \;\di\sigma_1(0)-\di\sigma_2(0)\\
\di\sigma_1(0)-\di\sigma_2(0)\; & \;\di\sigma_1(0)+\di\sigma_2(0)
\end{array}
\right)
\,,
\end{equation*}
which displays correlations between the COM and relative coordinates as long as the initial squeezing parameters of the two defect oscillators differ.

After turning on the coupling to the reservoir, the COM motion of the two defects thermalizes after a transient time $\tth<\trev$ which gives rise to the covariance matrix 
\begin{equation}
  \di\Sigma^\crc(\tth)=\left(
\begin{array}{cc}
\di\sigma^+(\di T) & 0\\
0 & \di\sigma^-(\tth)
\end{array}
\right)\,.
\label{eq:SigmaCOM_t}
\end{equation}
Here, the time-independent submatrix of the COM reads
\begin{equation}
\di\sigma^+(\di T)=
\left(
\begin{array}{cc}
\DXsquare & 0\\
0 & \DPsquare
\end{array}
\right)
\label{eq:SigmaPlusT}
\end{equation}
and contains the variances ${\DXsquare=\DXsquare(\di T)}$ and ${\DPsquare=\DPsquare(\di T)}$ on the diagonal. The actual values of ${\DXsquare}$ and ${\DPsquare}$ are numerically determined and depend on the initial temperature~$\di T$ of the reservoir. The time-dependent covariance matrix of the relative coordinate 
\begin{equation}
\di\sigma^-(\tth)=  T_-(\tth)\,\di\sigma^-(0)\, T^\tp_-(\tth)
\label{eq:SigmaMinusThermal}
\end{equation}
describes the free time evolution of the initial covariance matrix
\begin{equation}
\di\sigma^-(0)=\frac12\left(\di\sigma_1(0)+\di\sigma_2(0)\right)
\label{eq:InitialCovMatrixRelative}
\end{equation}
with the help of the orthogonal matrix 
\begin{equation}
T_-(\di t)=
\left(
\begin{array}{cc}
\phantom{-}\cos\di t & \sin\di t \\
-\sin\di t & \cos\di t
\end{array}
\right)
\label{eq:SymplecticT}
\end{equation}
that follows from the solution~\eqref{eq:relativeMotionModel1}. 
By transforming the covariance matrix~\eqref{eq:SigmaCOM_t} back to the original coordinates, we finally obtain the covariance matrix of the steady state,
\begin{equation*}
 \di\Sigma(\tth)=\frac12\left(
\begin{array}{cc}
\di\sigma^+(\di T)+\di\sigma^-(\tth) & \di\sigma^+(\di T)-\di\sigma^-(\tth)\\
\di\sigma^+(\di T)-\di\sigma^-(\tth) & \di\sigma^+(\di T)+\di\sigma^-(\tth)
\end{array}
\right)\,,
\end{equation*}
which now exhibits correlations between the first and second defect oscillator.

\subsection{Derivation of the auxiliary functions for the logarithmic negativity\label{App:LogNegFirstModel}}

In this section, we present the main steps of the derivation of the analytic expressions~\eqref{eq:DeterminantMod1} and~\eqref{eq:tildeDeltat}, which are used for the evaluation of the logarithmic negativity in Sec.~\ref{Sec:LogNegVariancesModel1}. We thereby take advantage of the determinant identity 
\begin{align}
\det(A\pm B)=&\det A+\det B 
\label{eq:DetAB}\\
&\pm \det\left(
\begin{array}{cc}
 a_{11} & a_{12} \\
 b_{21} & b_{22}
\end{array}
\right)
\pm \det \left(
\begin{array}{cc}
 b_{11} & b_{12} \\
 a_{21} & a_{22}
\end{array}
\right)\,.
\nonumber
\end{align}
which holds true for any two matrices~${A=(a_{ik})\in \mathds{C}^{2\times 2}}$ and~${B=(b_{ik})\in \mathds{C}^{2\times 2}}$.

In order to find the expression for the determinant~\eqref{eq:DeterminantMod1}, we first recall Eqs.~\eqref{eq:DeterminateCOM}, \eqref{eq:SigmaCOM_t},~\eqref{eq:SigmaPlusT} and~\eqref{eq:SigmaMinusThermal} to obtain
\begin{equation}
\det\di \Sigma(\tth)=\DXsquare\,\DPsquare\,\det\di \sigma^-(0)\,.
\label{eq:DetSigmaTth}
\end{equation}
Using the definition of~$\di \sigma^-(0)$, Eq.~\eqref{eq:InitialCovMatrixRelative}, and inserting the initial covariance matrices of the two oscillators~\eqref{eq:CovarianceSystemdimStandard}, we find with $\Delta\di \phi=\di\phi_2-\di\phi_1$
\begin{equation*}
\det\di \sigma^-(0)=\frac{1}{16}\det\left[S(e^{2\di r_1})+O^\tp(\Delta\di \phi)\,S(e^{2\di r_2})\,O(\Delta\di \phi)\right]\,.
\end{equation*}
The last expression can be easily evaluated with the help of identity~\eqref{eq:DetAB}, which yields after some minor algebra 
\begin{align}
\det\di \sigma^-(0)=& \,\frac18\Big(1+\cosh(2\di r_1)\cosh(2\di r_2)\Big.
\nonumber\\
& \Big. -\cos\left(\Delta\di\phi\right)\sinh(2\di r_1)\sinh(2\di r_2)\Big)\,.
\label{eq:DetSigmaMinus}
\end{align}
Substitution of the last expression into Eq.~\eqref{eq:DetSigmaTth} provides the expression~\eqref{eq:DeterminantMod1} for the determinant of $\di \Sigma(\tth)$.

To derive the time-dependent auxiliary function~\eqref{eq:tildeDeltat}, we start from Eqs.~\eqref{eq:tildeDeltaCOM}, \eqref{eq:SigmaCOM_t}, and~\eqref{eq:SigmaMinusThermal} and obtain with the orthogonality of $T_-(\di t)$
\begin{align}
\tilde \Delta(\tth)=&\det\di\sigma^+(\tth;\di T)+\det\di\sigma^-(0)
\nonumber\\
&-\det\left[\di\sigma^+(\tth;\di T)-\di\sigma^-(0)\right]\,,
\label{eq:tildeDeltatthcalc1}
\end{align}
where
\begin{equation*}
 \di\sigma^+(\tth;T)=T^\tp_-(\tth)\, \di\sigma^+(\di T)\, T_-(\tth)\,.
\end{equation*}
By applying the identity~\eqref{eq:DetAB} to Eq.~\eqref{eq:tildeDeltatthcalc1}, we find after a straightforward calculation 
\begin{align*}
\tilde \Delta(\tth)=&\,\frac12(\DXsquare+\DPsquare)\left(\di\sigma^-_{11}(0)+\di\sigma^-_{22}(0)\right)\\
&+\frac12(\DXsquare-\DPsquare)\left[(\di\sigma^-_{11}(0)-\di\sigma^-_{22}(0))\cos(2\tth)\right.\\
&+\left.2\,\di\sigma^-_{12}(0)\sin(2\tth)\right]\,.
\end{align*}
When we combine the two terms in the bracket of the last equation in a single cosine, we obtain the general form~\eqref{eq:tildeDeltat} of the auxiliary function. The resulting coefficients $\tilde \Delta_0$ and $\tilde \Delta_2$ can be rewritten in terms of the determinant and trace of $\di \sigma^-(0)$ according to
\begin{equation}
\tilde \Delta_0=\frac12\Big(\DXsquare+\DPsquare\Big)\tr{\di \sigma^-(0)}
\label{eq:DeltaZeroApp}
\end{equation}
and 
\begin{equation}
 \tilde \Delta_2=\frac12\Big|\DXsquare-\DPsquare\Big|\sqrt{(\tr{\di \sigma^-(0)})^2-4\det\di\sigma^-(0)}.
\label{eq:DeltaTwoApp}
\end{equation}
From the initial covariance matrices~\eqref{eq:CovarianceSystemdimStandard} and the definition~\eqref{eq:InitialCovMatrixRelative}, we obtain for the trace
\begin{equation*}
\tr{\di \sigma^-(0)}=\frac12\Big(\cosh(2\di r_1)+\cosh(2\di r_2)\Big)\,,
\end{equation*}
which together with the determinant~\eqref{eq:DetSigmaMinus} finally yields
\begin{align*}
(\tr{\di \sigma^-(0)})^2&\!-\!4\det\di\sigma^-(0)=\frac14\Big(\sinh^2(2\di r_1)+\sinh^2(2\di r_2)\Big.\\
\Big.&+2\cos(\Delta\di\phi)\,\sinh(2\di r_1)\sinh(2\di r_2)\Big)\,.
\end{align*}
Substitution of the last two expressions into Eqs.~\eqref{eq:DeltaZeroApp} and~\eqref{eq:DeltaTwoApp} finally concludes our derivation of the coefficients~\eqref{eq:DeltaZero} and~\eqref{eq:Deltatwo}.



\begin{thebibliography}{99}
\bibitem{Footnote:Rubin}
The microscopic model shown in Fig.~\ref{fig1} reduces to Rubin's one-dimensional crystal with one embedded defect when we choose the specific parameter values $\Omega=0$, $\omega_B=0$, $\gamma=\kappa/2$ and introduce the mass $M_0=4 M$ together with the new position and momentum operators ${X_0=X_+/2}$ and ${P_0=2 P_+}$ for the center-of-mass motion of the two oscillators.
\bibitem{footnote:residue}
This should be done by applying the inverse Laplace transformation to recover~$\di X_+(\di t)$ before performing the thermodynamic limit.
\bibitem{footnote:trev}
In our simulations, we determine the revival time by the equivalent expression 
$\trev\approx 2 \pi [\max\limits_j\,(\di\omega^+_{j+1}-\di\omega^+_j)]^{-1}\,.$
\bibitem{footnote:decay}
Note that a decrease of the mass ratio $\di m$ leads to an increase of the cutoff frequency and thus to a faster decay of the memory-friction kernel.
\end{thebibliography}
\end{document}